\documentclass[preprint*]{JHEP3} 

\usepackage{epsfig,multicol}

\newcommand\fverb{\setbox\pippobox=\hbox\bgroup\verb}
\newcommand\fverbdo{\egroup\medskip\noindent%
                        \fbox{\unhbox\pippobox}\ }
\newcommand\fverbit{\egroup\item[\fbox{\unhbox\pippobox}]}
\newbox\pippobox

\title{Studies of New Vector Resonances\\ 
at the CLIC Multi-TeV $e^+e^-$ Collider}

\author{Marco Battaglia\\
        CERN, CH-1211 Gen\`eve 23, Switzerland\\
        E-mail: \email{marco.battaglia@cern.ch}}

\author{Stefania De Curtis\\
I.N.F.N.,
Sezione di Firenze,  I-50019 Sesto F., Italy\\
E-mail: \email{decurtis@fi.infn.it}}

\author{Daniele Dominici\\
Dipartimento di Fisica, Universit\`a di Firenze,  I-50019 Sesto
F., Italy\\
I.N.F.N.,
Sezione di Firenze,  I-50019 Sesto F., Italy\\
E-mail: \email{dominici@fi.infn.it}}

\received{}             
\revised{}
\accepted{}             

\preprint{\hepph{0210351}}  

\abstract{Several models predict the existence of new vector resonances in the
multi-TeV region, which can be produced in high energy $e^+e^-$ collisions in the
s-channel. In this paper, we review  the existing limits on the masses of 
these resonances from {\sc Lep/Slc} and {\sc Tevatron} data and from atomic parity 
violation, in some specific models. We study the potential of a multi-TeV $e^+e^-$ 
collider, such as {\sc Clic}, for the determination of their properties and nature.}

\keywords{Beyond Standard Model, e+e- Experiments}

\begin{document}

\newcommand{\Zo}{\mathrm{Z}}
\newcommand{\Zp}{\mathrm{Z}'}
\newcommand{\LL}{\mathrm{LL}}
\newcommand{\LR}{\mathrm{LR}}
\newcommand{\RL}{\mathrm{RL}}
\newcommand{\RR}{\mathrm{RR}}
\newcommand{\Ll}{\mathrm{L}}
\newcommand{\Rr}{\mathrm{R}}
\newcommand{\eemm}{\mathrm{e}^+\mathrm{e}^- \rightarrow \mu^+ \mu^-}
\newcommand{\mumu}{\mu^+ \mu^-}
\newcommand{\tautau}{\tau^+ \tau^-}
\newcommand{\bb}{\mathrm{b \bar{b}}}
\newcommand{\cc}{\mathrm{c \bar{c}}}
\newcommand{\bea}{\begin{eqnarray}}
\newcommand{\eea}{\end{eqnarray}}
\newcommand{\be}{\begin{equation}}
\newcommand{\ee}{\end{equation}}
\newcommand{\nn}{\nonumber}
\def\f{\frac}
\def\gt{\tilde g}
\def\ct{{\tilde c}_\theta }
\def\st{{\tilde s}_\theta }
\def\sb{s_\beta }
\def\cb{ c_\beta }
\def\eps{\epsilon}

\def\s{s_\theta }
\def\c{c_\theta }

\section{Introduction}
While the core of the physics program of a TeV-class linear collider (LC) can be 
already largely defined on the basis of what we know today, the signals from new physics 
which could be probed by a multi-TeV collider, such as {\sc Clic}~\cite{clic} at
1~TeV$<\sqrt{s}<$~5~TeV, belong to a significantly broader domain. Still, one of the 
most striking manifestation of new physics will come from the sudden increase of the 
$e^+e^- \rightarrow f \bar f$ cross section indicating the s-channel production of a new
particle. There are several theories which predict the existence of such a resonance. 
In this paper we study the sensitivity of {\sc Clic} to scenarios including new 
vector boson resonances.
A first class consists of models with extra gauge bosons such as a new neutral $Z'$ 
gauge boson. This is common to both GUT-inspired $E_6$ models and to Left-Right (LR) 
symmetric models. They are discussed in Section~2. Models of dynamical electroweak 
symmetry breaking also predict the existence of new resonances in the TeV region. In 
particular, we consider the degenerate BESS (D-BESS) model, which describes a pair of 
narrow and nearly degenerate vector and  axial-vector states~\cite{dbess}, in Section~3.
Additional resonances are also introduced by recent theories of gravity with extra 
dimensions in the form of Kaluza-Klein graviton and gauge boson excitations. A five 
dimensional extension of the Standard Model (SM) is discussed in Section 4. Beyond 
discovery, it will be essential to accurately measure their masses,
widths, production and decay properties to determine their nature and identify which 
kind of new physics they manifest. The recently proposed little Higgs models, a new 
approach to the hierarchy problem, predict also new vector bosons in the TeV range, see 
for instance~\cite{Wacker:2002ar}. Their possible signature at a multi-TeV collider
deserves further study.
{\sc Clic} will also be sensitive to new vector gauge 
bosons at mass scales much beyond the kinematic threshold. In Section 5 we discuss the 
statistical accuracy for the determination of the cross sections, $\sigma_{f \bar f}$, 
and forward-backward asymmetries, $A_{FB}^{f\bar f}$, for $\mu^+\mu^-$,  $b \bar b$ and 
$t \bar t$ at $\sqrt{s}$ = 3~TeV. These accuracies 
will be used to establish the sensitivity reach to indirect effects of new vectors as
$Z'$ gauge bosons  and Kaluza-Klein excitations of the photon and of the $Z^0$ boson.

\section{$Z'$ Boson studies}

One of the simplest extensions of the SM is to introduce an
additional $U(1)$  gauge symmetry, whose breaking scale is close
to the Fermi scale. This extra symmetry is predicted in some grand
unified theories and in other models. For example, in $E_6$ scenarios
we have the following additional $U(1)$ current
 \be
 J_{Z^\prime\mu}^f = J_{\chi\mu}^f {\cos\theta_6}+
J_{\psi\mu}^f {\sin\theta_6}
 \ee
with different models  parameterised by specific values of the
angle $\theta_6$. The $\chi, \psi$ and $\eta$ models
correspond to the values $\theta_6$=0, $\theta_6=\pi/2$ and
$\theta_6=-\tan^{-1}\sqrt{5/3}$ respectively.
In the LR models, the new $Z_{LR}$ boson couples to the current
 \be
 J_{Z^\prime\mu}=\alpha_{LR}J_{3R\mu}-\frac{1}{2\alpha_{LR}}
J_{(B-L)\mu} \ee with $\alpha_{LR}=\sqrt{g_R^2/g_L^2
\cot^2\theta_W-1}$.
The vector and axial-vector couplings of the $Z'$ boson to the SM fermions, for 
$E_6$-inspired and for LR models, are given in Table \ref{coupzp}, 
assuming:
 \be
 J_{Z^\prime\mu}^f = \bar f\left[\gamma_\mu {v_f^\prime}+\gamma_\mu \gamma_5
{a_f^\prime}\right] f
 \ee
and the parametrisation $\theta_2 = \theta_6 + \tan^{-1} \sqrt{5/3}$.
\TABLE[t]{
\begin{tabular}{|c|c|}
\hline
 Extra-U(1) & LR $(g_L=g_R)$ \\
\hline\hline
 $v_e^\prime=-\frac 1 4 s_\theta\left( c_2+\sqrt{\frac 5 3}
s_2\right)$ &$  v_e^\prime=\left(-\frac 1 4 +
 s_\theta^2\right)/\sqrt{c_{2\theta}}$\\
\hline
 $a_e^\prime=\frac 1 4 s_\theta\left(-\frac 1 3
c_2+\sqrt{\frac 5 3} s_2\right)$&$a_e^\prime=-\frac 1 4
\sqrt{c_{2\theta}}$\\
\hline $ v_u^\prime=0 $&$ v_u^\prime=\left(\frac 1 4 - \frac 2 3
s_\theta^2\right)/\sqrt{c_{2\theta}}$\\
\hline
 $a_u^\prime=- \frac 1 3 s_\theta c_2 $ &$  a_u^\prime=\frac 1 4
\sqrt{c_{2\theta}}$\\
\hline
 $v_d^\prime=\frac 1 4 s_\theta\left( c_2+\sqrt{\frac 5 3}
s_2\right)$ &$  v_d^\prime=\left(-\frac 1 4 +
\frac 1 3 s_\theta^2\right)/\sqrt{c_{2\theta}}$\\
\hline
 $a_d^\prime=a_e^\prime$&$a_d^\prime=a_e^\prime  $ \\
\hline
\end{tabular}
\caption
{Vector and axial-vector couplings for
the $E_6$-inspired and the LR models, $s_{\theta} = \sin \theta$, 
$s_2=\sin\theta_2$, $c_2=\cos\theta_2$, $c_{2\theta} = \cos 2 \theta$ 
with $\theta_2=\theta_6+\tan^{-1}\sqrt{5/3}$ and $\theta\equiv
\theta_W$.}
\label{coupzp}
}
Finally, a useful reference is represented by the so-called sequential standard model
(SSM), which introduces an extra $Z'$ boson with the same couplings of the SM $Z^0$ 
boson.

There exist several constraints on the properties of new neutral vector gauge bosons.
Direct searches for a new $Z'$ boson also set lower limits on the 
masses~\cite{Abe:1997fd,kobel}. These are summarised in Table~\ref{tab:2} for various 
models. An extra $Z'$ naturally mixes with the SM $Z^0$ boson. The present precision 
electroweak data constrain  the mixing angle, $\theta_M$, within a few mrad, and the 
masses as shown in Table~\ref{tab:2}~\cite{Langacker:2001ij,lep}. 
\TABLE[t]{
\begin{tabular}{|c|c|c|c|c|c|}
\hline
&$\chi$ & $\psi$ &$\eta$& $LR$ & SSM \\
\hline
{\sc Cdf}&595 & 590 & 620 & 630 & 690\\ {\sc Lep}&673 & 481
& 434 & 804 & 1787\\
\hline
\end{tabular}
\caption{95\%~C.L.\ limit on $M_{Z'}$ (GeV) from $\sigma(pp\to Z')
B(Z'\to ll)$ ({\sc Cdf} data) and  from the  average of the four 
{\sc Lep} experiments, for the mixing angle $\theta_M=0$.}
\label{tab:2}
}
A third class of constraints is derived from atomic parity violation (APV) data.
We update here the bounds obtained in~\cite{Casalbuoni:1999yy} and based on the
1999 result of weak charge $Q_W$ in the Cesium atomic parity 
experiment~\cite{Bennett:1999pd}, 
which indicated a $\simeq 2.6 \sigma$ discrepancy w.r.t. the SM prediction. A series of 
theoretical papers have since improved the prediction of $Q_W$, by including the effect 
of the Breit interaction among electrons \cite{breit} and by refining the
calculation of radiative corrections \cite{radapv}. A complete re-analysis of 
the parity non conserving amplitude for the $6S\to 7S$ transition in Cesium has been 
performed~\cite{dzuba}, which improves on the theoretical uncertainties. In addition 
the self-energy and vertex QED radiative corrections have been shown to yield a large 
negative contribution to the parity non conserving amplitude~\cite{Kuchiev:2002fg},
bringing the result on the extraction of $Q_W$ from the Cesium data to:

\be
Q_W=-72.71\pm0.29_{\rm exp}\pm0.39_{\rm theor}
\label{qw}
\ee

The corresponding SM prediction, obtained for $m_t$=175.3$\pm$4.4~GeV and 
$M_H$=(98$^{+51}_{-35}$)~GeV, is $Q_W^{({\rm SM})}=-73.10\pm 0.03$\cite{PDG}. 
Here we inflate the uncertainty to $\pm 0.13$ to account for the hadronic-loop and 
other uncertainties. The new result given in eqs. (\ref{qw}) agrees well with the SM 
prediction. Models involving extra neutral vector bosons can modify the $Q_W$ value 
significantly. Assuming no $Z^0-Z'$ mixing, the contribution to the weak charge due to 
the direct exchange of the $Z'$ is given by
 \be
\delta_N Q_W=16 a_e^\prime [(2 Z+N)v_u^\prime +(Z+2N)v_d^\prime]
\frac {M_{Z}^2}{M_{Z^\prime}^2}
 \ee
where $Z=55$, $N=78$ for Cs and $a_f,v_f,a'_f,v'_f$ are the $Z^0$ and $Z'$ couplings
to fermions (see Table~\ref{coupzp}). For $E_6$-inspired  models, bounds on $M_{Z'}$ 
are derived as function of the angle $\theta_6$, by comparing the predicted value for
the weak charge with that in eq.~(\ref{qw}) (see Figure~\ref{zpp}). The lower limits on
$M_{Z'}$ are less stringent than, or comparable to those in 
Table~\ref{tab:2}. However, as these 
bounds are very sensitive to the actual value of $Q_W$ and its uncertainties, further 
determinations may improve the situation.

\FIGURE[t]{\epsfig{file=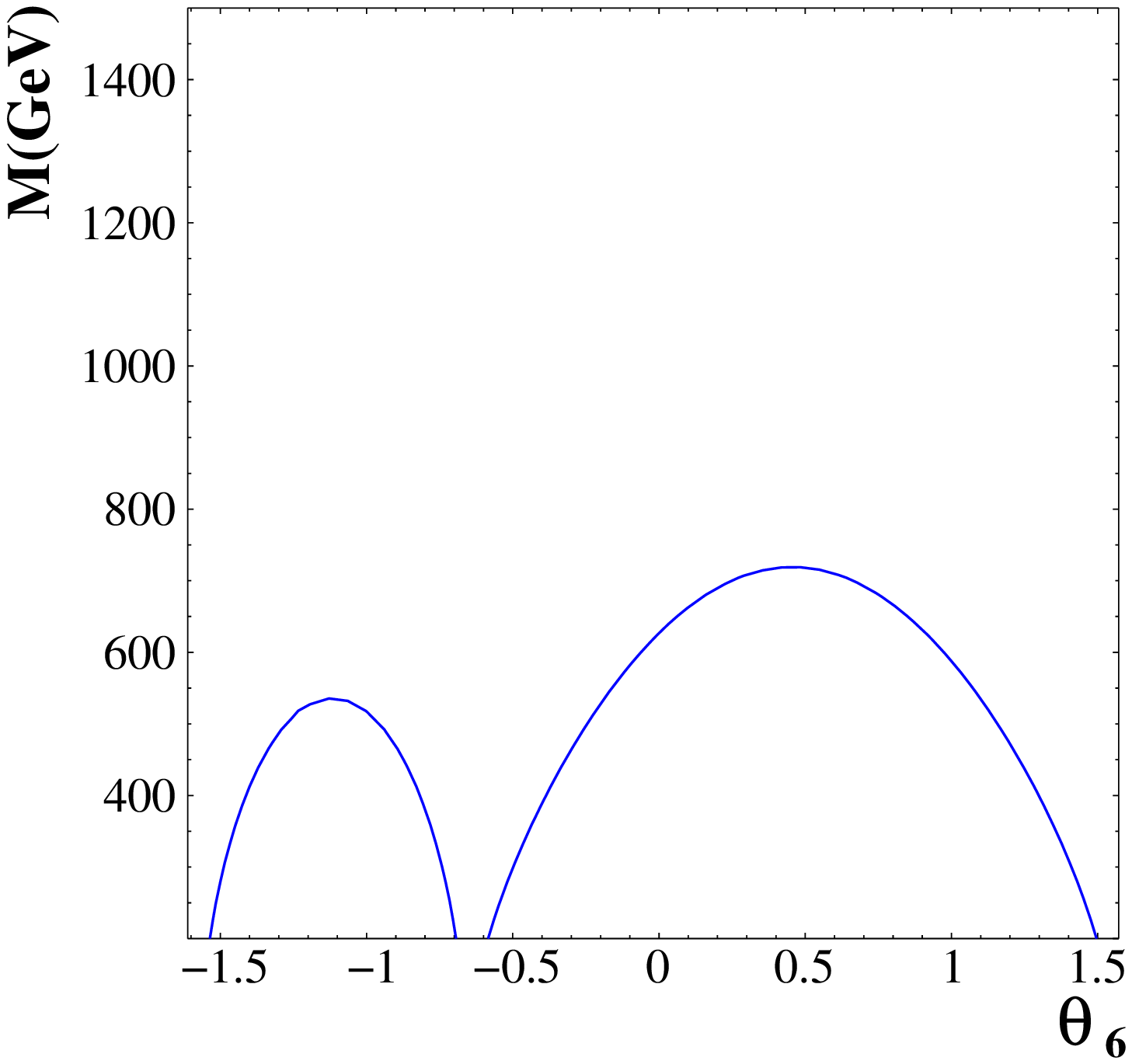,width=9.cm,height=6.5cm,clip}
\caption{95\%~C.L.\
lower bounds on $M_{Z'}$ at fixed  $\theta_6$ from the
Cesium atomic parity experiment result given in eq.
(\ref{qw}). }
\label{zpp}}

In the case of the LR model, neglecting mixing and ${W'}$ contributions, we get 
$\delta_N Q_W=-(M^2_{Z}/M^2_{Z'}) Q_W^{(SM)}$  \cite{Casalbuoni:1999yy}, corresponding
to the 95\%~C.L.\ bound, $ M_{Z_{LR}}> 665$ GeV.

Finally, for the SSM $Z'$ boson we get a contribution $\delta_N
Q_W=(M^2_{Z}/M^2_{Z'}) Q_W^{(SM)}$  \cite{Casalbuoni:1999yy}, leading to the
95\%~C.L.\ bound, $M_{Z'_{SSM}}>1010$ GeV.

The {\sc Lhc} hadron collider will push the direct sensitivity to new vector gauge 
bosons beyond the TeV threshold. With an integrated luminosity of 100~fb$^{-1}$, 
{\sc Atlas} and {\sc Cms} are expected to observe signals from $Z'$ bosons for masses 
up to 4-5~TeV depending on the specific model~\cite{Godfrey:2002tn}.

Extra-$U(1)$ models can be accurately tested at a future linear $e^+e^-$ collider, 
operating in the multi-TeV region, such as {\sc Clic}. With an expected 
effective production cross section $\sigma(e^+e^- \to Z'_{SSM})$ of $\simeq$~15~pb,
including the effects of ISR and luminosity spectrum, a $Z'$ resonance will tower over 
a $q \bar q$ continuum background of $\simeq$~0.13~pb. While the observation of such
signal is granted, the accuracy that can be reached in the study of its properties 
depends on the quality of the accelerator beam energy spectrum and on the detector
response, including accelerator induced backgrounds. One of the main characteristics
of the {\sc Clic} collider is the large design luminosity, 
$L=10^{35}$~cm$^{-2}$~s$^{-1}$ at $\sqrt{s}$ = 3~TeV for its baseline parameters,
obtained in a regime of strong beamstrahlung effects. 
The optimisation of the total luminosity and its fraction in the peak has been studied 
for the case of a resonance scan. The {\sc Clic} luminosity spectrum has
been obtained with a dedicated beam simulation program~\cite{Schulte:2001kh} for the 
nominal parameters at $\sqrt{s}$ = 3~TeV. In order to study the systematics from
the knowledge of this spectrum, the modified Yokoya-Chen parametrisation~\cite{peskin} 
has been adopted. In this formulation, the beam energy spectrum is described in terms of
$N_\gamma$, the number of photons radiated per $e^{\pm}$ in the bunch, the beam energy 
spread in the linac $\sigma_p$ and the fraction $\cal{F}$ of events outside the 0.5\% of
the centre-of-mass energy. Two sets of parameters have been considered, obtained by 
modifying the beam size at the interaction point and therefore the total luminosity and 
its fraction in the highest energy region of the spectrum: CLIC.01 with
${\cal{L}}$=1.05$\times10^{35}$~cm$^{-2}$ s$^{-1}$ and $N_{\gamma}$=2.2 and CLIC.02 with
${\cal{L}}$=0.40$\times10^{35}$~cm$^{-2}$ s$^{-1}$ and $N_{\gamma}$=1.2. 
The $Z'$ mass and width
can be determined by performing either an energy scan, like the $Z^0$ line-shape scan
performed at {\sc Lep}/{\sc Slc}, and also foreseen for the $t \bar t$ threshold, or an 
auto-scan, by tuning the collision energy just above the top of the resonance and 
profiting of the long tail of the luminosity spectrum to probe the resonance peak. For 
the first method both di-jet and di-lepton final states can be considered, while for the
auto-scan only $\mu^+ \mu^-$ final states can provide with the necessary accuracy for 
the $Z'$ energy. $e^+e^- \rightarrow Z'$ events have been generated for $M_{Z'}$ = 3~TeV,
including the effects of ISR, luminosity spectrum and $\gamma \gamma$ backgrounds, 
assuming SM-like couplings, corresponding to a total width 
$\Gamma_{Z'_{SSM}} \simeq$ 90~GeV. The resonance widths for 
extra-$U(1)$ models as well as for other SM extensions with additional vector bosons are
shown in Figure~\ref{figwidth} as a function of the relevant model parameters. 
\FIGURE[t]{
\epsfig{file=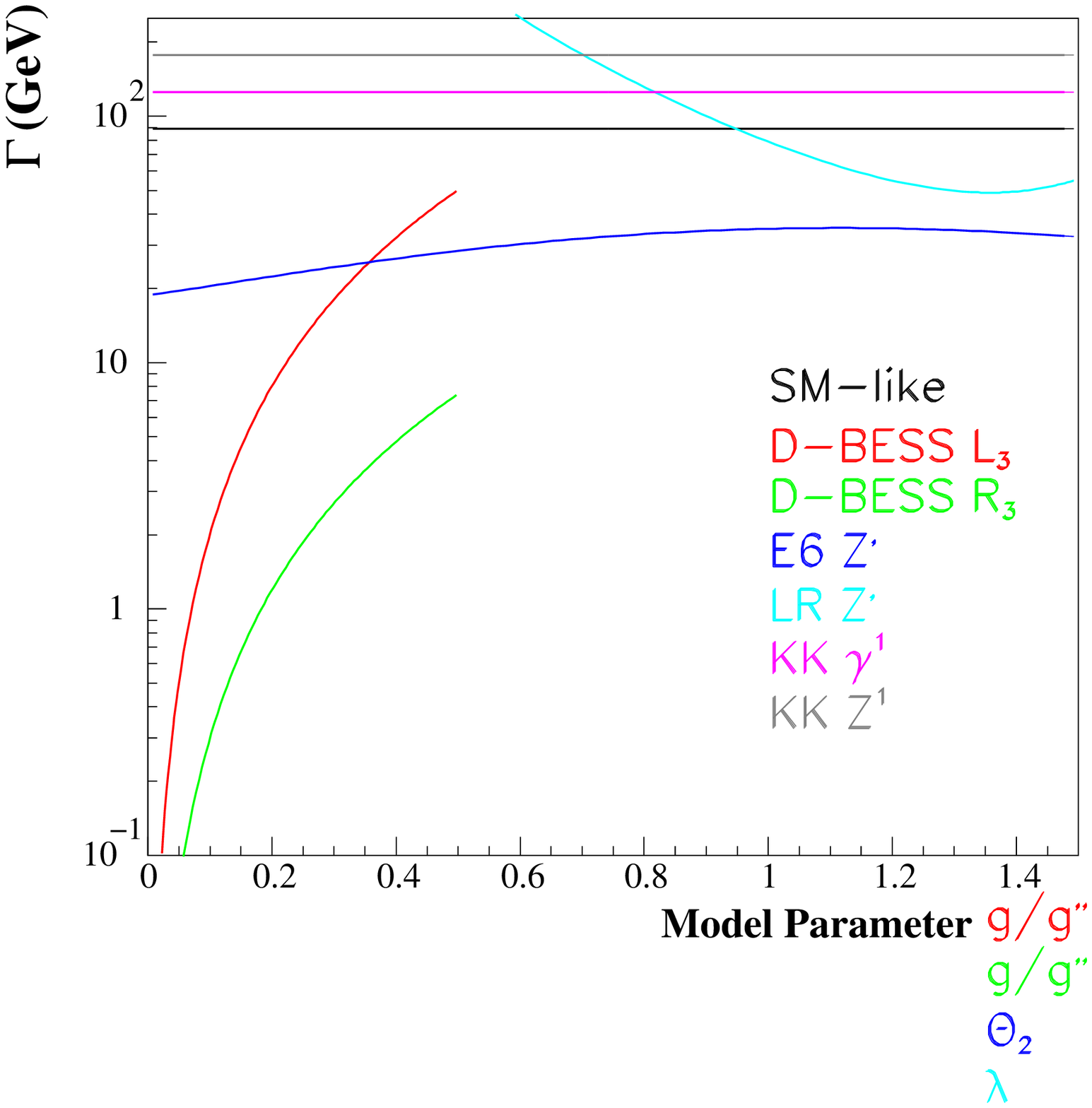,width=10.0cm,height=9.0cm,clip}
\caption[]{Widths of new gauge vector bosons as a function of the relevant parameters: 
$\theta_2$ for $Z'_{E_6}$, $\lambda=g_L/g_R$ for $Z'_{LR}$~\cite{alta}, $g/g''$ for
D-BESS. The KK $Z^{(1)}$ width has a negligible dependence on the mixing 
angle $\sin\beta$. The $Z'_{E_6}$ and $Z'_{LR}$ widths are computed by assuming only 
decays into SM fermions.}
}\label{figwidth}

\FIGURE[t]{
\epsfig{file=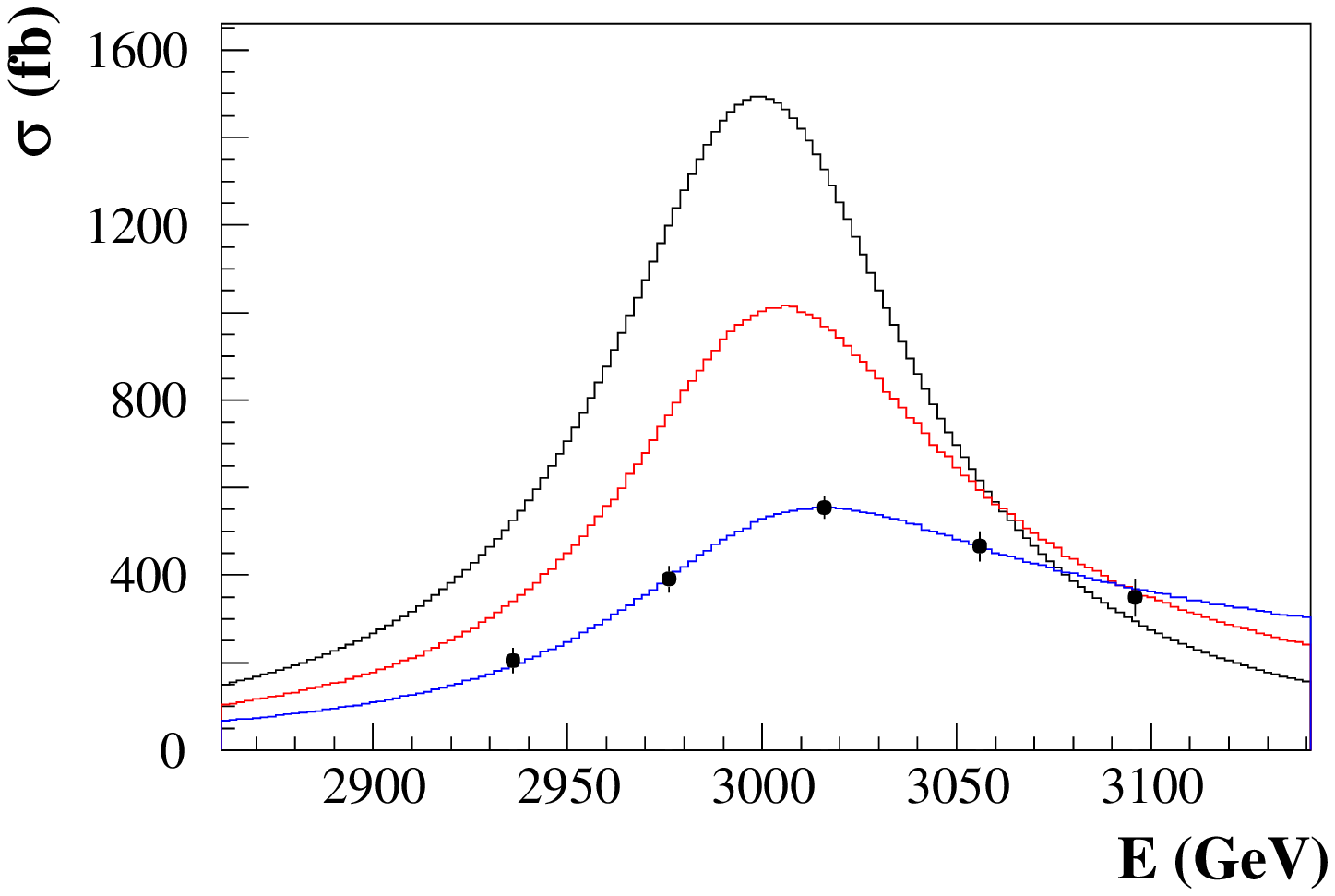,width=11.5cm,height=7.5cm,clip}
\caption{$Z'_{SSM} \to \ell^+ \ell^-$ resonance profile obtained by
energy scan. The Born production cross-section, the cross
section with ISR included and that accounting for the {\sc Clic} 
luminosity spectrum (CLIC.01) and tagging criteria are shown.}
\label{fig:zpxs}}
A data set of  1~ab$^{-1}$ has been assumed for the CLIC.01 beam
parameters and of 0.4~ab$^{-1}$ for CLIC.02, corresponding to one year
(10$^{7}$~s) of operation at nominal luminosity. This has been shared
in a five point scan (see Figure~\ref{fig:zpxs}) 
and $M_{Z'}$, $\Gamma_{Z'}/\Gamma_{Z^0}$
and $\sigma_{peak}$ have been extracted from a $\chi^2$ fit to the
predicted cross section behaviour for different mass and width values 
(see Table~\ref{tab:3})~\cite{Battaglia:2001fr}. 
The dilution of the analysing power due to the beam energy
spread is appreciable, as can be seen by comparing the statistical
accuracy from a fit to the pure Born cross section after including
ISR and beamstrahlung effects. Still, the relative statistical
accuracies are better than 10$^{-4}$ on the mass and $5 \times
10^{-3}$ on the width. In the case of wide resonances, there is an advantage 
in employing the broader luminosity spectrum, CLIC.01, which offers 
larger luminosity. Sources of systematics from the knowledge of
the shape of the luminosity spectrum have also been estimated. In
order to keep $\sigma_{syst} \le \sigma_{stat}$ it is necessary to
control $N_{\gamma}$ to better than 5\% and the fraction $\cal{F}$ of
collisions at $\sqrt{s} < 0.995 \sqrt{s_{0}}$ to about 1\%~\cite{Battaglia:2001dg}.

\TABLE[t]{
\begin{tabular}{|l|c|c|c|}
\hline
Observable & Breit Wigner & CLIC.01 & CLIC.02 \\ \hline
$M_{Z'}$ (GeV) & 3000 $\pm$ .12  & $\pm$ .15 &  $\pm$ .21 \\
$\Gamma_{Z'}/\Gamma_{Z^0}$ & 1. $\pm$ .001 & $\pm$ .003  & $\pm$ .004 \\
$\sigma^{eff}_{peak}$ (fb) & 1493 $\pm$ 2.0 & 564 $\pm$ 1.7 & 669 $\pm$ 2.9 \\
\hline
\end{tabular}
\caption{Results of the fits for the cross section scan of a $Z'_{SSM}$ obtained 
by assuming no radiation and ISR with the effects of two different choices of the 
{\sc Clic} luminosity spectrum.}
\label{tab:3}
}

\section{Study of the D-BESS model}

Present precise electroweak data are consistent with the realization of the Higgs 
mechanism with a light elementary Higgs boson. But as the Higgs boson has so far 
eluded the direct searches, it remains important to assess the sensitivity of 
future colliders to strong electroweak symmetry breaking (SSB) scenarios. SSB models 
are based on low energy effective Lagrangians which provide a phenomenological 
description of the Goldstone boson dynamics. Possible new vector resonances produced 
by the strong interaction responsible for the electroweak symmetry breaking can be 
introduced in this formalism as gauge bosons of a hidden symmetry. A description of a 
new triplet of vector resonances is obtained by considering an effective Lagrangian 
based on the symmetry $SU(2)_L\otimes SU(2)_R\otimes SU(2)_{local}$~\cite{bess}. 
The new vector fields are a gauge triplet of the $SU(2)_{local}$. They acquire mass as 
the $W^{\pm}$ and the $Z^0$ bosons. By enlarging the symmetry group of the model, 
additional vector and axial-vector resonances can be introduced.

The degenerate BESS model (D-BESS)~\cite{dbess} is a realization
of dynamical electroweak symmetry breaking with decoupling. The
D-BESS model introduces two new triplets of gauge bosons, which
are almost degenerate in mass, ($L^\pm$, $L_3$), ($R^\pm$, $R_3$).
The extra parameters  are a new gauge coupling constant $g''$ and
a mass parameter $M$, related to the scale of the underlying
symmetry breaking sector. In the charged sector the $R^\pm$ fields
are not mixed and $M_{R^\pm}=M$, while $M_{{L}^\pm}\simeq M
(1+x^2)$ for small $x=g/g''$ with $g$  the usual $SU(2)_W$ gauge
coupling constant. The $L_3$, $R_3$ masses are given by
$M_{L_3}\simeq  M\left(1+ x^2\right),~~ M_{R_3}\simeq M \left(1+
x^2 \tan^2 \theta\right)$ where $\tan \theta = g'/g$ and $g'$ is the usual 
$U(1)_Y$ gauge coupling constant. 
These resonances are narrow (see Figure~\ref{figwidth}) 
and almost degenerate in mass with 
$\Gamma_{L_3}/M\simeq 0.068~ x^2$ and $\Gamma_{R_3}/M\simeq 0.01~
x^2$, while the neutral mass splitting is: $\Delta
M/M=(M_{L_3}-M_{R_3})/M \simeq \left( 1-\tan^2 \theta \right)
x^2\simeq 0.70 ~x^2$. This model respects the present bounds from electroweak 
precision data since the $S,T,U$ (or $\epsilon_1, \epsilon_2, \epsilon_3$) 
parameters vanish at the leading order in the limit of large $M$ due to an 
additional custodial symmetry. Therefore, electroweak data set only loose 
bounds on the parameter space of the model. We have studied these bounds by 
considering the latest experimental values of the $\epsilon$ parameters coming 
from the high energy data \cite{epsi}: 
\be \epsilon_1= (5.4\pm
1.0)\cdot 10^{-3},~~ \epsilon_2= (-9.7\pm 1.2)\cdot 10^{-3},~~
\epsilon_3= (5.4\pm 0.9)\cdot 10^{-3} \label{epsiexp} 
\ee \\
We have included radiative corrections, taken to be the same as in the SM, with 
the Higgs mass as a cut-off \cite{dbess}. For $m_t=175.3$~GeV and
$m_H=1000$~GeV  one has \cite{epsilon}: $\eps_1^{\rm rad}=
3.78\cdot 10^{-3}$, $\eps_2^{\rm rad}= -6.66\cdot 10^{-3}$,
$\eps_3^{\rm rad}=6.65\cdot 10^{-3}$. The 95\%~C.L.\ bounds on the
parameters of the D-BESS model are shown in Figure \ref{dom:fig1}.
Comparable bounds come from the direct search at the {\sc
Tevatron}~\cite{dbess}.

The {\sc Lhc} can discover these new resonances,
which are produced through a $q \bar q$ annihilation through
their leptonic decay $q{\bar q'}\to L^\pm,W^\pm\to (e \nu_e)
\mu\nu_\mu$ and $q{\bar q}\to L_3,R_3,Z,\gamma\to
(e^+e^-)\mu^+\mu^-$.
\TABLE[t]{
\begin{tabular}{|c |c |c |c |c |c|}
\hline
$g/g''$ & $M$ & $\Gamma_{L_3}$ & $\Gamma_{R_3}$ & $S/\sqrt{S+B}$ & $\Delta M$
\\
& (GeV) &(GeV) & (GeV) & {\sc Lhc} ($e+\mu$) & {\sc Clic}
\\
 \hline  0.1 &
1000 & 0.7 & 0.1 &17.3 &
\\
0.2 & 1000 & 2.8 & 0.4 & 44.7 &
\\\hline
0.1 & 2000 & 1.4 & 0.2 &3.7 &
\\
0.2 & 2000 & 5.6 & 0.8 & 8.8&
 \\\hline
0.1 & 3000 & 2.0 & 0.3 &(3.4)& 23.20 $\pm$ .06
\\
0.2 & 3000 & 8.2 & 1.2 &(6.6)& 83.50 $\pm$ .02
 \\
\hline
\end{tabular}
\caption{Sensitivity to production of the $L_3$ and $R_3$ D-BESS resonances at the 
{\sc Lhc} for ${\cal{L}}=$100(500)~fb$^{-1}$ with $M=$1,2(3)~TeV and accuracy on the 
mass splitting at {\sc Clic} for ${\cal{L}}=$1~ab$^{-1}$.} 
\label{tab:4}}
The relevant observables are the di-lepton transverse and invariant masses.
The main backgrounds, left to these channels after the lepton isolation
cuts, are the Drell-Yan processes with SM gauge bosons
exchange in the electron and muon channel. The study has been performed using
a parametric detector simulation~\cite{redi}. Results are given in
Table~\ref{tab:4} for the combined electron and muon channels
for ${\cal{L}}=100$~fb$^{-1}$, except for $M=$~3~TeV where 500~fb$^{-1}$ are assumed.

The discovery limit at {\sc Lhc}, with ${\cal{L}}=100$~fb$^{-1}$, is $M\sim
2$~TeV for $g/g''=0.1$. Beyond discovery, the possibility to disentangle 
the characteristic double peak structure depends strongly on $g/g''$ and
smoothly on the mass. 

The LC can also probe this multi-TeV region
through the virtual effects in the cross-sections for $e^+e^-\to
{L_3},{R_3},Z,\gamma \to f \bar f $, at centre-of-mass energies below the 
resonances. Due to the presence of new spin-one resonances the annihilation channel 
in $f \bar f$ and $W^+W^-$ has a better sensitivity than the fusion channel.
\FIGURE[t]{
\begin{tabular}{l  r}
\includegraphics[width=0.46\textwidth,height=0.46\textwidth]{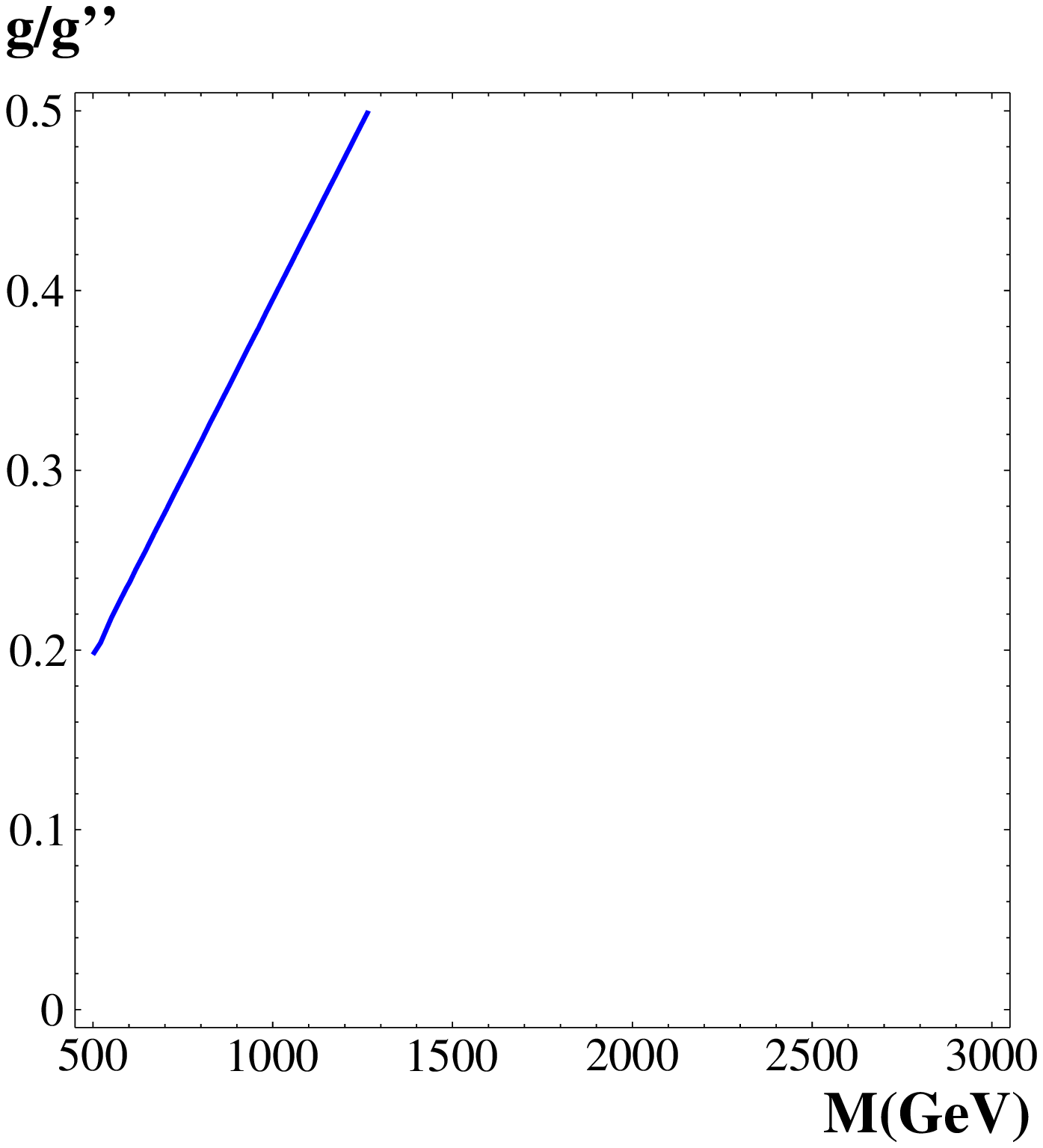}&
\includegraphics[width=0.45\textwidth,height=0.45\textwidth]{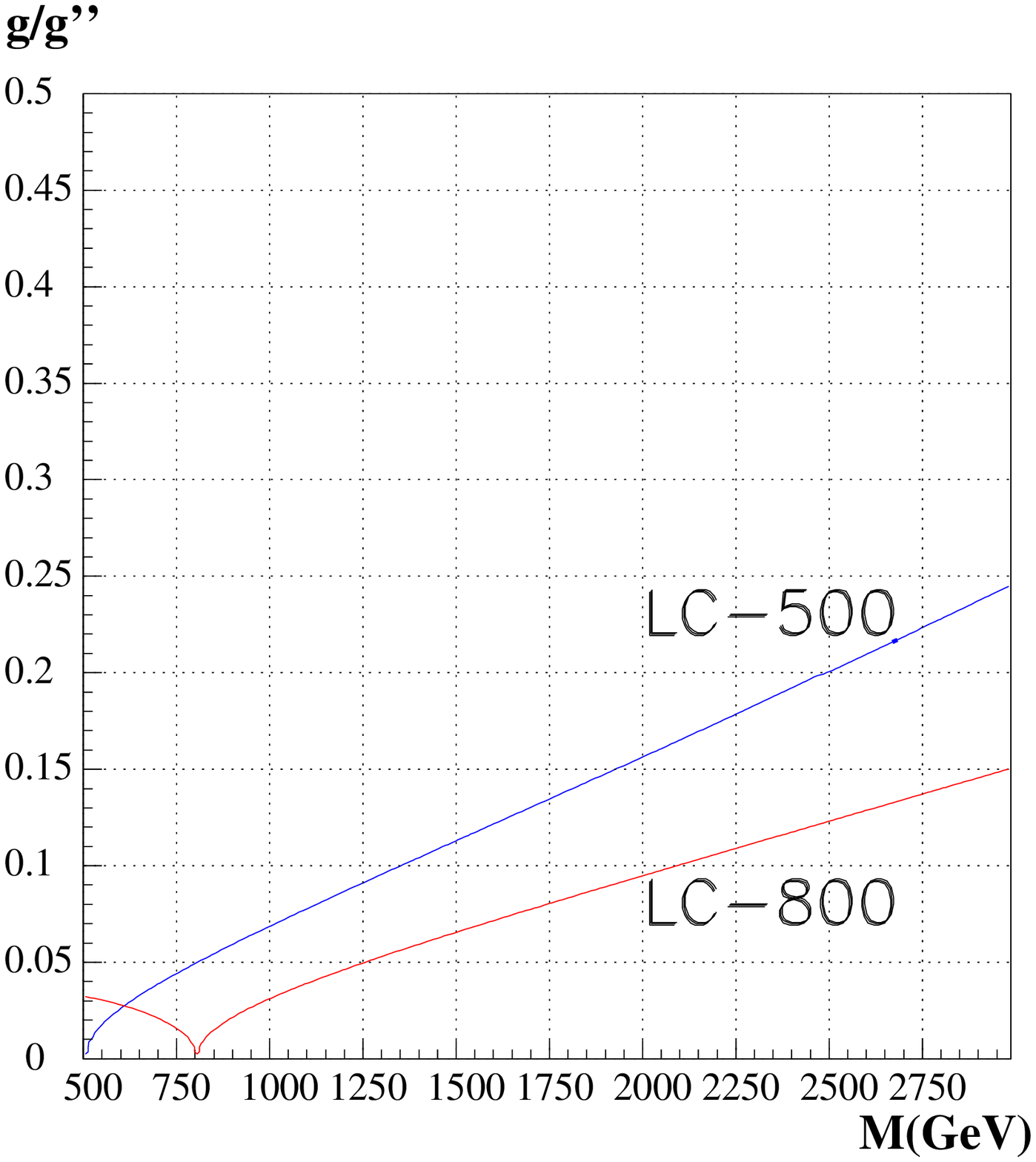}\\
\end{tabular}
\caption {95\%~C.L.\ contours in the plane
$(M,~ g/g'')$ from the present $\epsilon$ measurements (left-hand side)
and from measurements of $\sigma_{\mu^+\mu^-}$, $\sigma_{b \bar
b}$, $A_{FB}^{\mu\mu}$, $A_{FB}^{bb}$ at $e^+e^-$ linear colliders
with $\sqrt{s}=500(800)$~GeV  and ${\cal{L}}=1~$ab$^{-1}$ (right-hand
side). The allowed regions are below the curves.}
\label{dom:fig1}}
\FIGURE[t]{
\epsfig{file=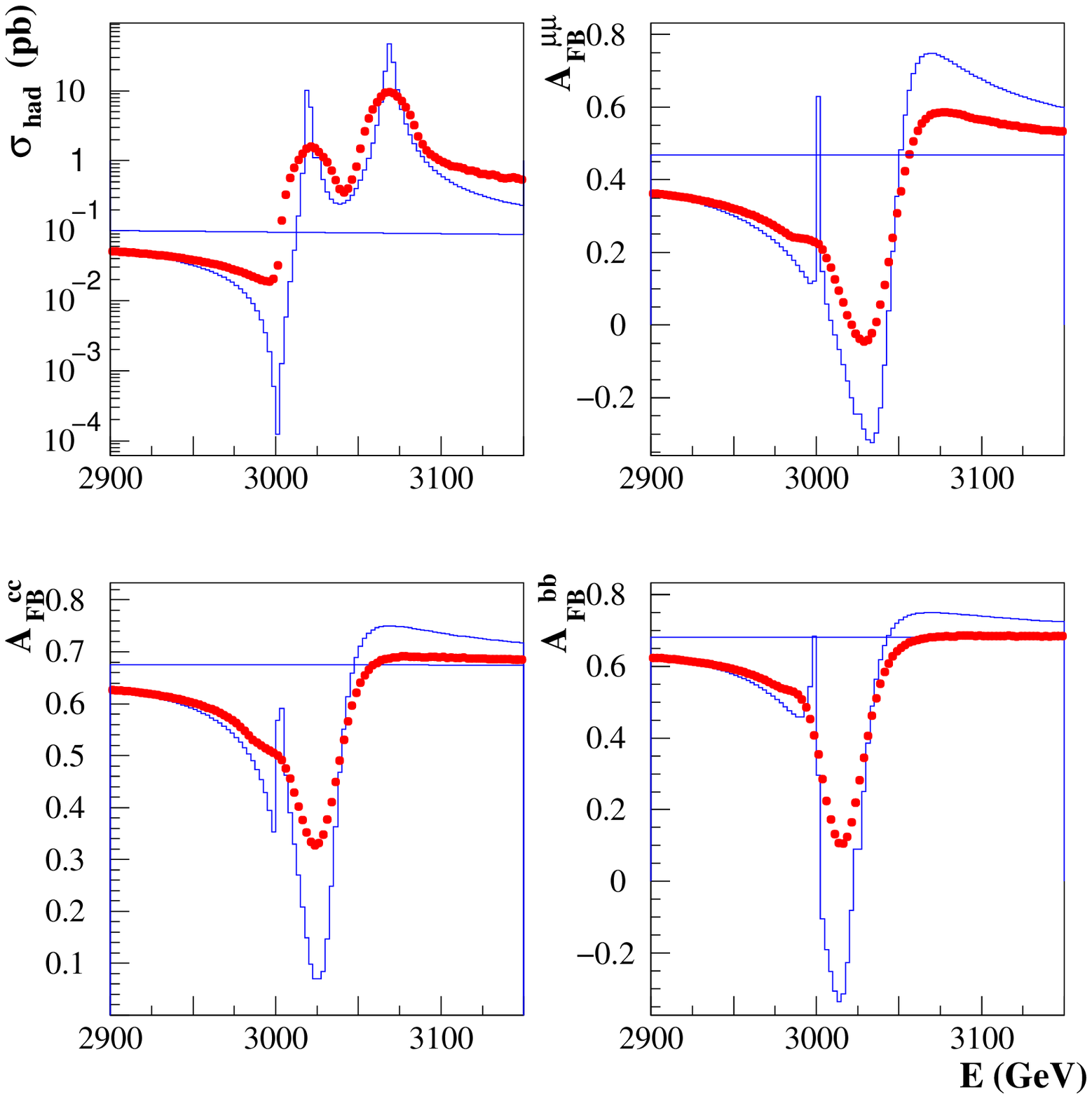,width=13.0cm,height=10.0cm,clip}
\vspace*{-0.25cm}
\caption{Hadronic cross section (upper left) and $\mu^+\mu^-$ (upper
right), $c \bar c$ (lower left) and $b \bar b$ (lower right)
forward-backward asymmetries at energies around 3~TeV. The
continuous lines represent the predictions for the D-BESS
model with $M$ = 3~TeV and $g/g''=0.15$, the flat lines the
SM expectation and the dots the observable D-BESS signal
after accounting for the CLIC.02 luminosity spectrum.}
\label{fig:afb}}
In the case of the D-BESS model, the $L_3$ and $R_3$ states are not strongly
coupled to $W$ pairs, making the $f\bar f$ final states the most favourable
channel for discovery. Analysis at $\sqrt{s}=$ 500 GeV and $\sqrt{s}=$ 800 GeV
is based on $\sigma_{\mu^+\mu^-}$, $\sigma_{b \bar b}$,
$A_{FB}^{\mu\mu}$ and $A_{FB}^{bb}$. We assume identification efficiencies of 
$\epsilon_\mu=95~\%$ and $\epsilon_b=60\%$ and  systematic uncertainties of
$\Delta \epsilon_\mu/\epsilon_\mu=0.5\%$,  $\Delta\epsilon_b/\epsilon_b=1\%$
as.

The sensitivity contours obtained for ${\cal{L}}=$ 1~ab$^{-1}$ are shown in 
Figure~\ref{dom:fig1}. The 3~TeV LC indirect reach is lower or 
comparable to that of the  {\sc Lhc}. However, the QCD background rejection 
essential for the {\sc Lhc} sensitivity still needs to be validated using full 
detector simulation and pile-up effects.


Assuming a resonant signal to be seen at the {\sc Lhc} or indirect evidence to be 
obtained at a lower energy LC, {\sc Clic} could measure the width and mass of this state
and also probe its almost degenerate structure~\cite{Casalbuoni:1999mm}. 
This needs to be validated when taking the luminosity spectrum and accelerator 
induced backgrounds into account. The ability to identify the model distinctive 
features has been studied using the production cross section and the flavour dependent
forward-backward asymmetries, for different values of $g/g''$. The resulting 
distributions are shown in Figure~\ref{fig:afb} for the case of the narrower 
CLIC.02 beam parameters. A characteristic feature of the cross section distributions is
the presence of a narrow dip, due to the interference 
of the $L_3$, $R_3$ resonances with the $\gamma$ and $Z^0$ and to cancellations of the 
$L_3$, $R_3$ contributions. Similar considerations hold for the asymmetries. In the 
case shown in Figure~\ref{fig:afb}, the effect is still visible after accounting for the
luminosity spectrum. In this analysis, the beam energy spread sets the main limit to 
smallest mass splitting observable. With realistic assumptions and 1~ab$^{-1}$ of data 
{\sc Clic} will be able to resolve the two narrow resonances for values of the coupling 
ratio $g/g''>$~0.08, corresponding to a mass splitting $\Delta M$ = 13~GeV for
$M=$ 3~TeV, and to determine $\Delta M$ with a statistical accuracy better than 
100~MeV (see Table~\ref{tab:4}).

\section{Kaluza-Klein excitations in theories with Extra-Dimensions}

Theories of quantum gravity have considered the existence of
extra-dimensions for achieving the unification of gravity at a
scale close to that of electroweak symmetry breaking. String
theories have recently suggested that the SM could live on a
$3+\delta$ brane with $\delta$ compactified large dimensions while
gravity lives on the entire ten dimensional bulk. The
corresponding models lead to new signatures for future colliders
ranging from Kaluza-Klein (KK) excitations of the
gravitons~\cite{HLZ} to KK excitations of the SM gauge fields with
masses in the TeV range~\cite{antoniadis}.

Among the models with extra dimensions we consider here a five-dimensional 
extension of the SM with fermions on the boundary. This predicts KK excitations 
of the SM gauge bosons with fermion couplings $\sqrt{2}$ larger compared to those 
of the SM \cite{Pomarol:1998sd}.
Masses of KK excitations of $W$, $Z^0$ and $\gamma$ are given by $M_n \simeq nM$, 
for large value of the fifth dimension compactification scale, $M$.

Indirect limits from electroweak measurements already exist and are
derived by considering the modifications in the electroweak observables 
at the $Z^0$ peak and at low energy \cite{ewlimits,Casalbuoni:1999ns}.
We discuss here the bounds derived from recent determination of the $\epsilon$ 
parameters, given on eq.~(\ref{epsiexp}), and from the APV results discussed in 
Section~2.

The contribution of the KK excitations of $W^{\pm}$, $Z^0$ and $\gamma$ to the $\epsilon$
parameters is given by:
\be
\eps_{1N}= -\c^2 X [ 1 + \sb^2 \f {\s^2}{\c^2} (1+ \cb^2)],~~
\eps_{2N}= -\c^2 X,~~
\eps_{3N}=- 2 \c^2 \sb^2 X\label{eps}
\ee
where $X=\pi^2 M_Z^2/(3M^2)$, the effective $\theta$ angle is defined through 
$ {G_F}/\sqrt{2}=
 {e^2}/({8 \s^2\c^2 M_Z^2})$
and $\beta$ parametrises the mixing between the KK excitations and the
SM gauge bosons~\cite{Casalbuoni:1999ns}.
Contributions from radiative corrections are included assuming SM values.
Additional radiative correction terms could originate from the additional 
charged and neutral Higgs bosons. These are not included here. 

The $95\%$~C.L.\ lower bounds on the scale $M$ at fixed $\sb$, coming from the $\eps$
observables, are given by the upper curves  in Figures \ref{boKK}.  
The two curves correspond to $m_t=175.3$ GeV and $m_H=98$ GeV
and $m_H=180$ GeV. We have used ~\cite{epsilon}: 
$\eps_1^{\rm rad}(98)=5.66\cdot 10^{-3}$, $\eps_2^{\rm rad}(98)= 
-7.56\cdot 10^{-3}$,
$\eps_3^{\rm rad}(98)=5.07\cdot 10^{-3}$ and
$\eps_1^{\rm rad}(180)=5.39\cdot 10^{-3}$, $\eps_2^{\rm rad}(180)= 
-7.41\cdot 10^{-3}$,
$\eps_3^{\rm rad}(180)=5.49\cdot 10^{-3}$

Bounds can be also obtained from low energy neutral current experiments. 
An effective current-current interaction Lagrangian was derived 
in~\cite{Casalbuoni:1999ns}. From this expression we compute the
relevant low energy observables. The atomic weak
charge $Q_W$ is given by \be Q_W= {Q}_{W}^{\mathrm{ (SM)}} [1 + \s^2 X
(\sb^2-1)^2]-4\frac{\s^2\c^2} {c_{2\theta}}{\rm Z}\Delta
\label{qwkk} \ee where ${Q}_W^{\mathrm{ (SM)}}$ has the SM expression for
$Q_W$ , Z is the atomic number and $\Delta = \c^2 X ( 1 - 2 \sb^2
- \sb^4 {\s^2}/ {\c^2})$. By using the determinations for the weak
charge of the Cs nucleus,  given in eq. (\ref{qw}), and its SM
prediction of $-73.10\pm0.13$, as in Section 2, we derive 95\%~C.L.\ bounds 
which results significantly below the corresponding high energy limits and are 
shown by the lower curves in Figure~\ref{boKK}.

Non observation of deviations in lepton pair production at {\sc Lhc}
can set limits on the compactification scale $M$. For example by considering
an effective luminosity of 5 fb$^{-1}$ one gets a bound of 
$M=6.7$~TeV~\cite{Antoniadis:1999bq}.

\FIGURE[t]{\epsfig{file=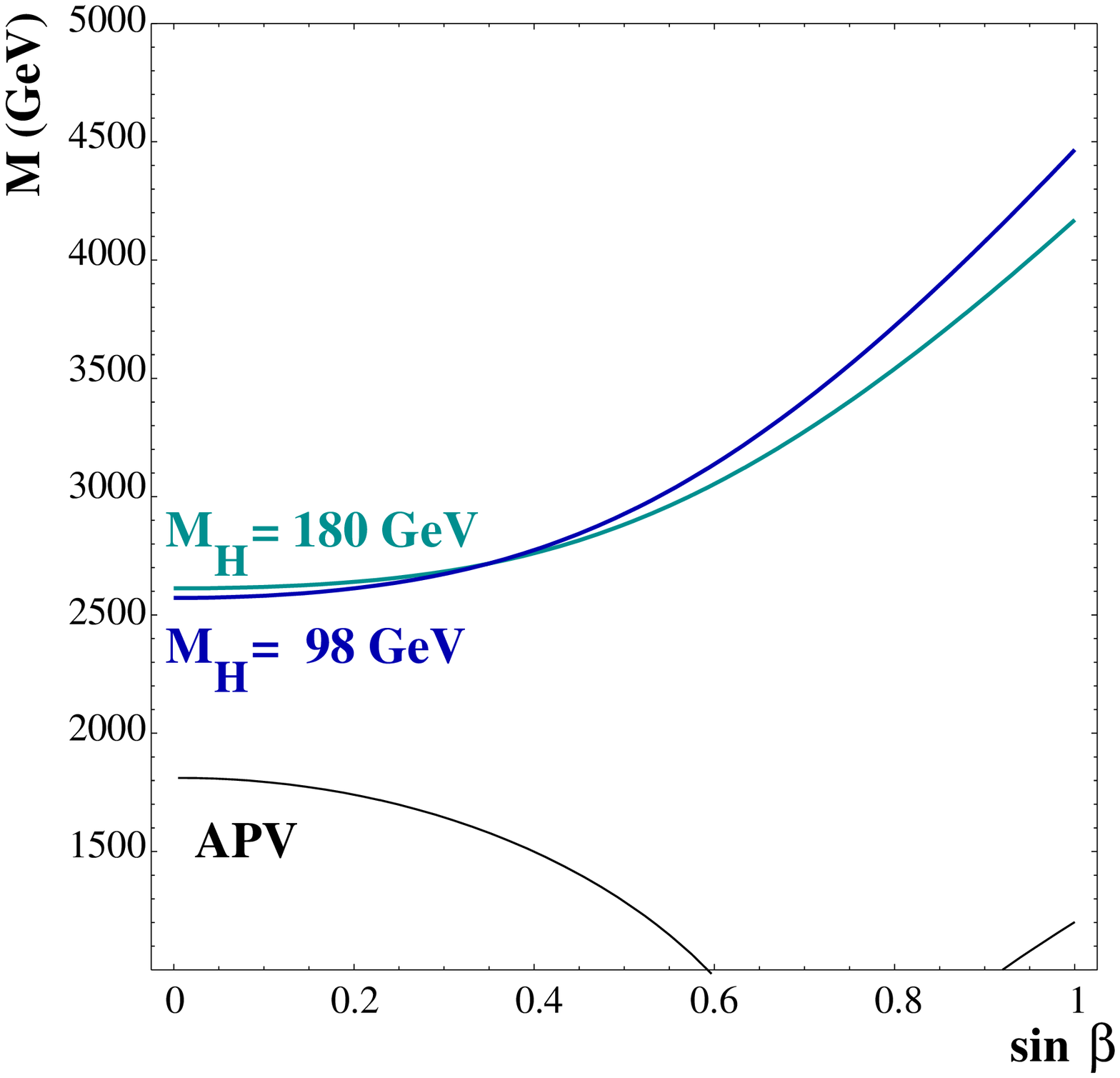,height=7.0cm,width=9.0cm,clip}
\caption{95\%~C.L.\ lower bounds on the compactification scale $M$, as function of 
$\sin\beta$, from the high energy precision measurements ($\epsilon$ parameters), 
for $m_t=175.3~GeV$ and two different values of $m_H$ and from the APV data. 
The regions below the lines are excluded.} 
\label{boKK}}

At {\sc Clic}, the lowest excitations $Z^{(1)}$ and $\gamma^{(1)}$ could be 
directly produced. Their widths are shown in Figure~\ref{figwidth}. 
Since the KK excitations of the photon do not mix with the other gauge vectors,
the $\gamma^{(1)}$ width does not depend on $\beta$. The $Z^{(1)}$ 
width has only a small correction, $\frac{\delta\Gamma_{Z^{(1)}}}{\Gamma_{Z^{(1)}}}=2 
\sin^2\beta \frac {m_Z^2}{M^2}$ which is not visible in the Figure.

Results for the $\mu^+\mu^-$ cross sections and forward-backward asymmetries at the 
Born level and after folding the effects of the CLIC.02 beam spectrum are shown in 
Figure~\ref{kksmear}. For comparison, we also present curves corresponding to the case
where only the $Z^{(1)}$ excitation is present.

\FIGURE[t]{
\begin{tabular}{l  r}
\includegraphics[width=0.45\textwidth,height=0.4\textwidth]{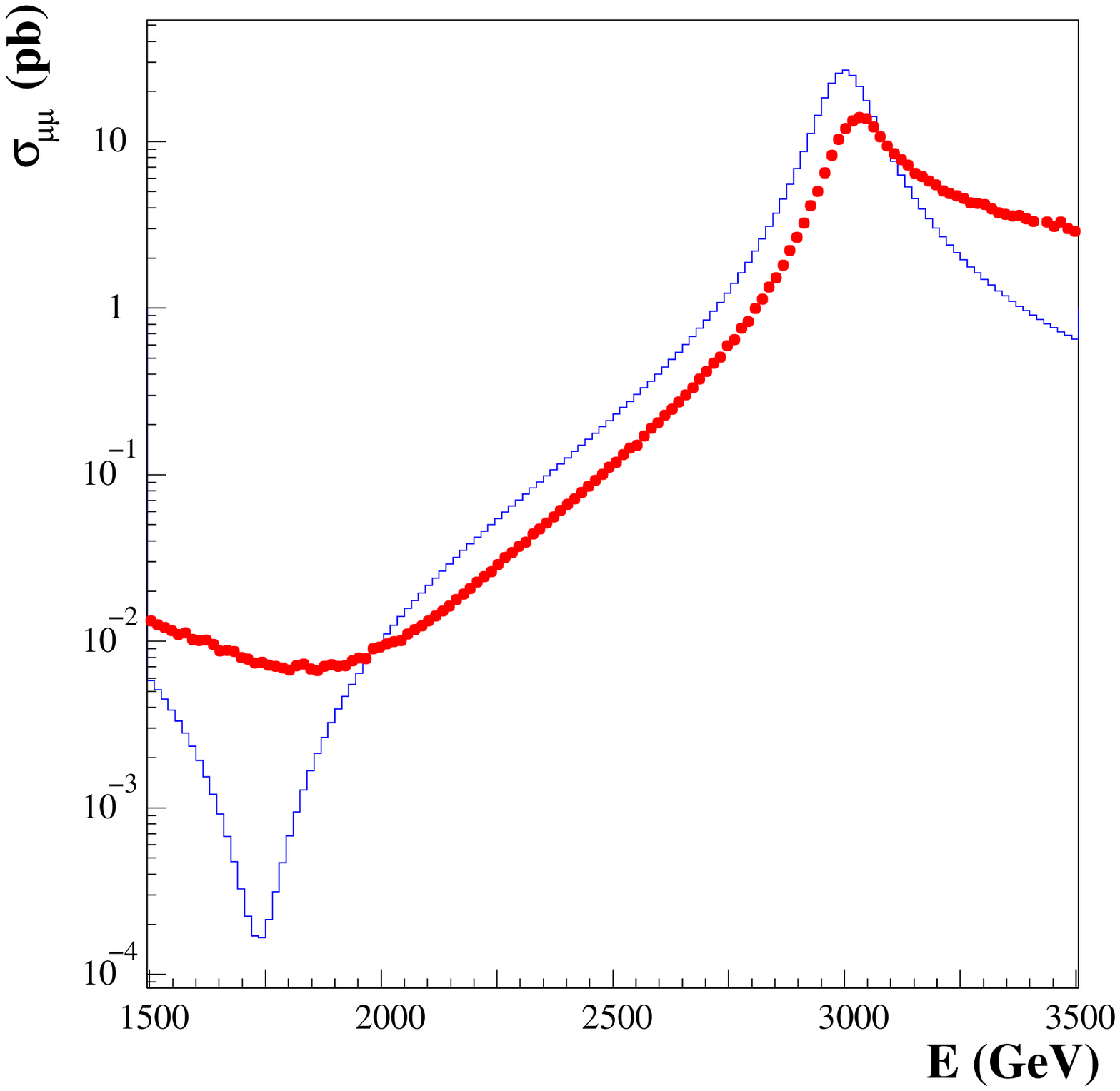}&
\includegraphics[width=0.45\textwidth,height=0.4\textwidth]{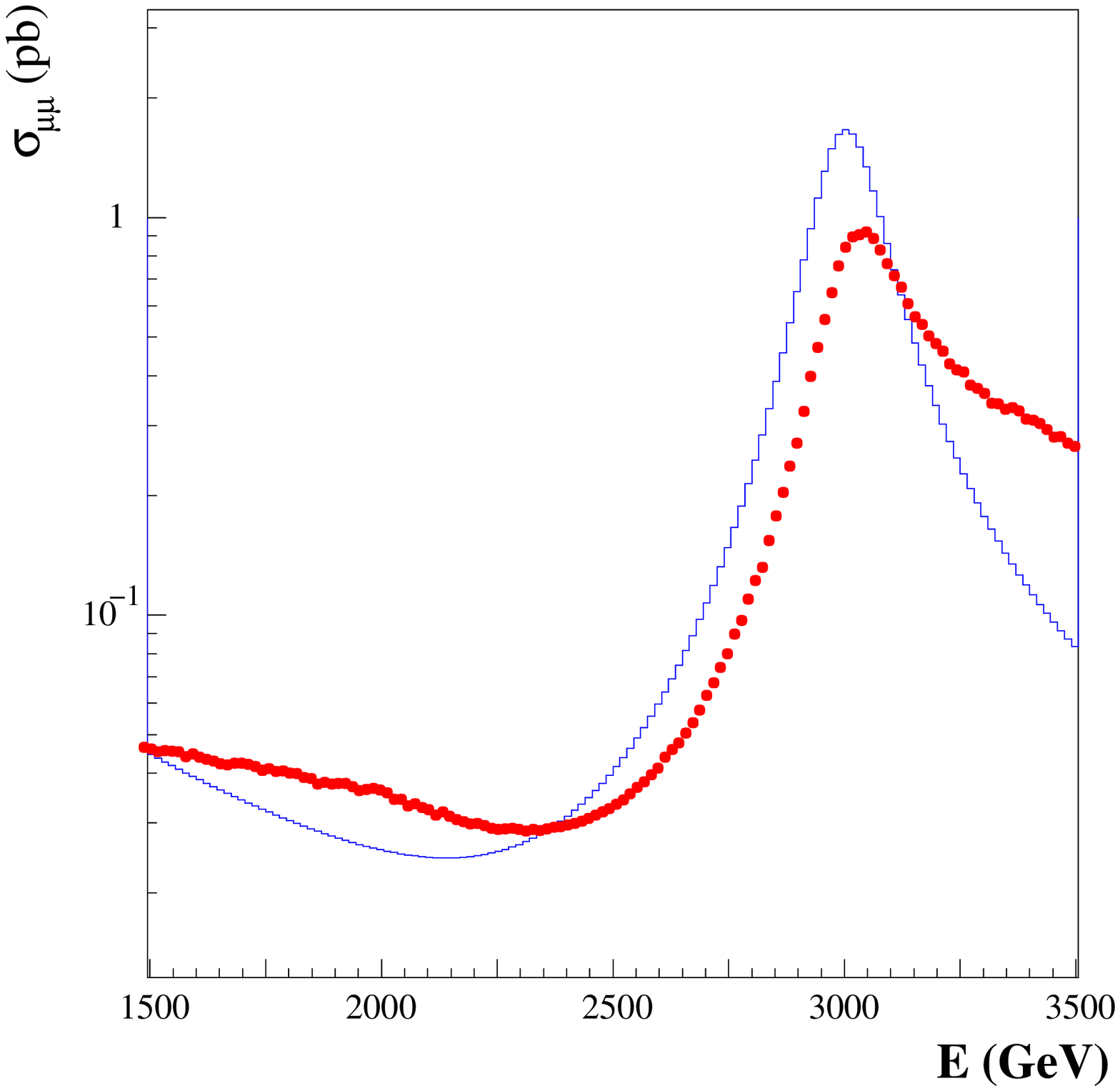}\\
\includegraphics[width=0.45\textwidth,height=0.4\textwidth]{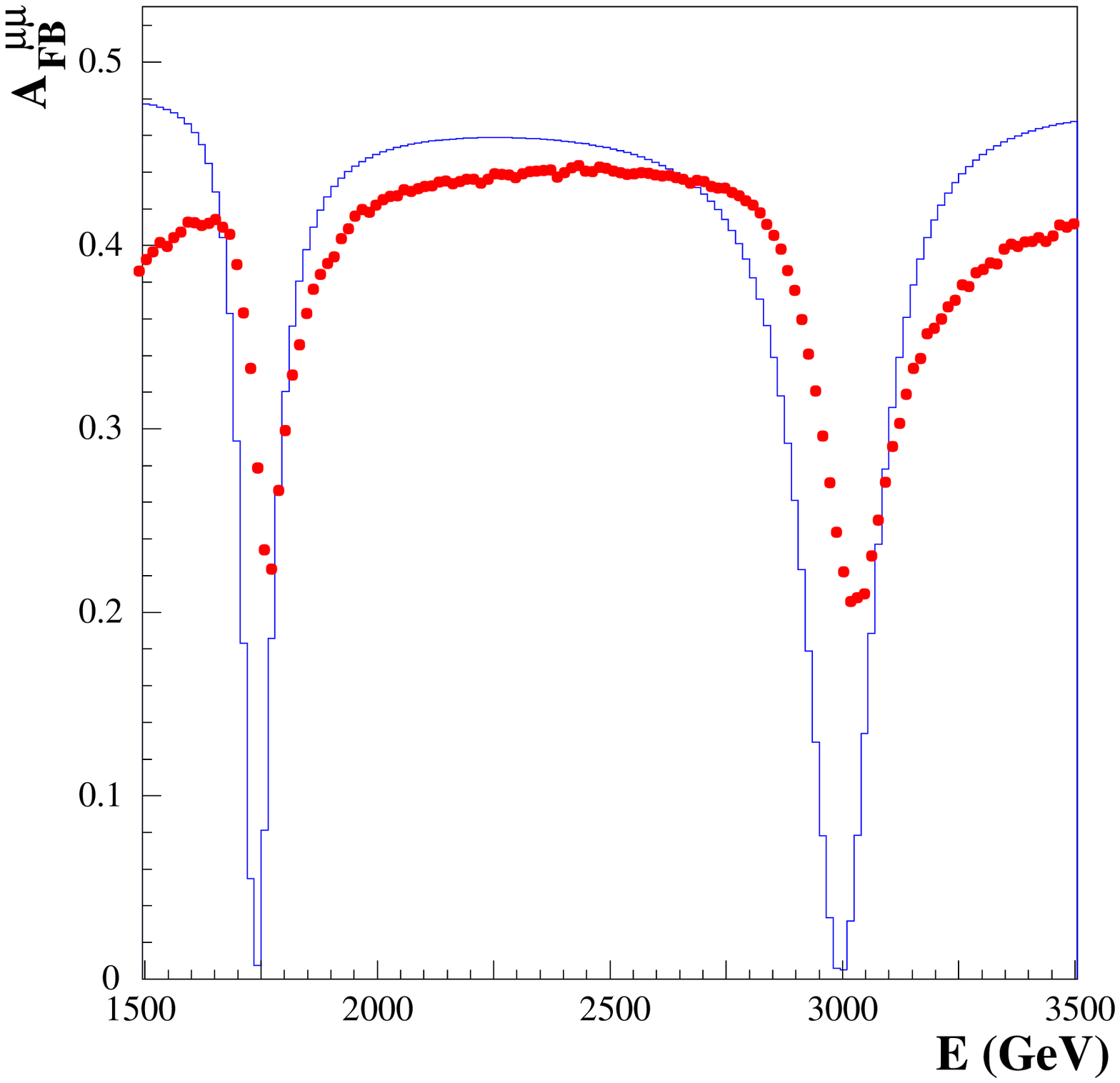}&
\includegraphics[width=0.45\textwidth,height=0.4\textwidth]{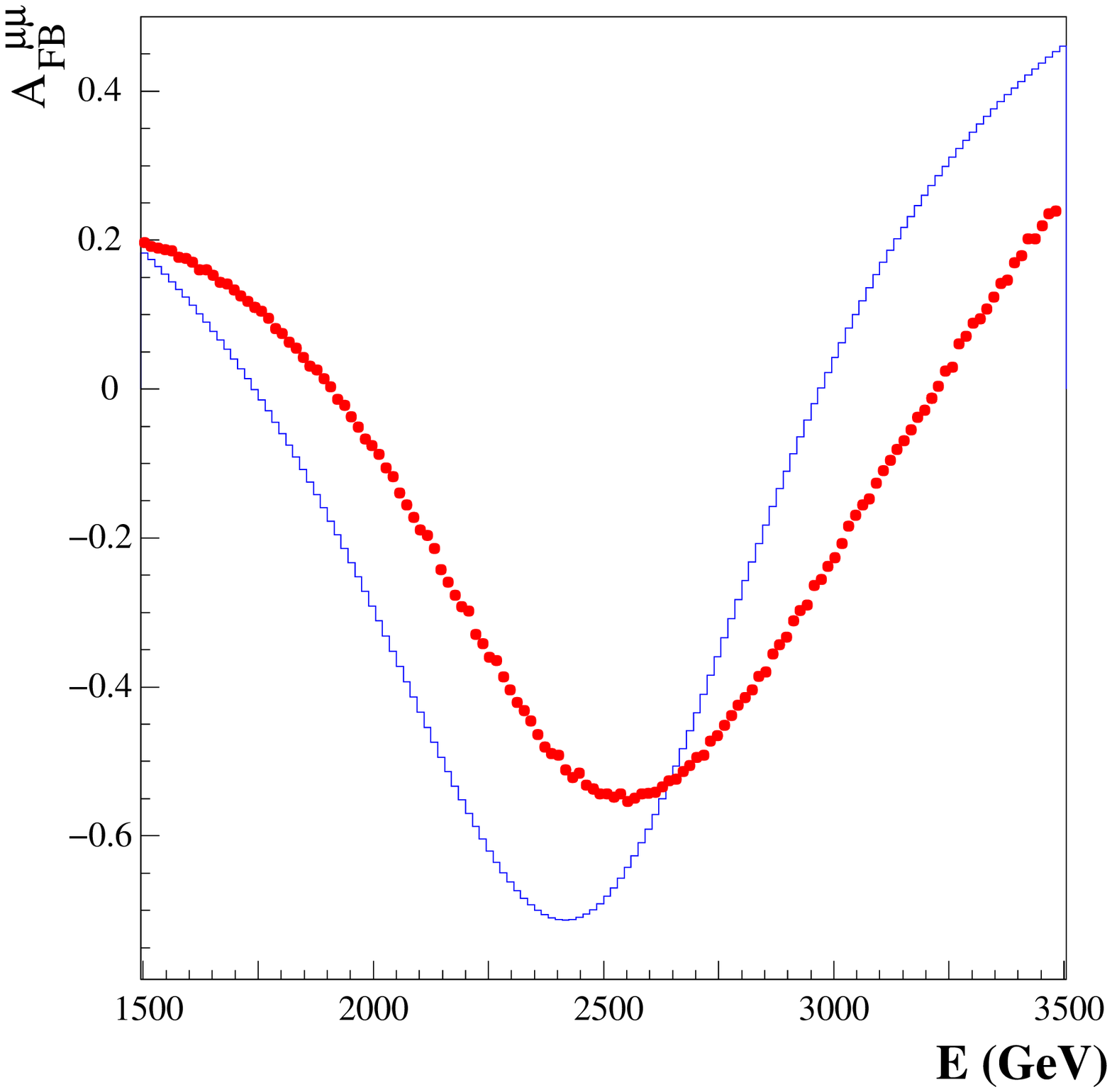}\\
\end{tabular}
\caption{$\mu^+\mu^-$ production cross sections and forward-backward asymmetries
in the 5D~SM including the lower KK excitations of the $Z^0$ and $\gamma$  with 
$M_{Z^{(1)}}\sim M_{\gamma^{(1)}} = 3~TeV$ (left) and in presence of only the $Z^{(1)}$ 
excitation (right). The continuous lines represent the Born-level expectations while 
the dots include the effect of the {\sc Clic} luminosity spectrum.}
\label{kksmear}}

\section{Electroweak observables 
and sensitivity to $Z'$ Boson and KK excitations}

The design energy of the {\sc Clic} collider generally matches the {\sc Lhc} 
sensitivity to new gauge vector bosons, allowing to systematically study their 
properties after their initial observation at the {\sc Cern} hadron collider. 

Precision electroweak measurements performed in multi-TeV $e^+e^-$ collisions can 
push the mass scales sensitivity, for these new phenomena, beyond the 10~TeV frontier.
We consider here the $\mu^+\mu^-$, $b \bar{b}$ and $t \bar{t}$ production cross sections
$\sigma_{f \bar{f}}$ and forward-backward asymmetries $A_{FB}^{f \bar{f}}$.  At the 
{\sc Clic} design centre-of-mass energies, the relevant $e^+e^- \rightarrow f \bar f$ 
cross sections are significantly reduced and the experimental conditions at the 
interaction region need to be taken into account in validating the accuracies on
electro-weak observables. Since the two-fermion cross section is of the order of only 
10~fb, it is imperative to operate the collider at high luminosity. This can be 
achieved only in a regime where beam-beam effects are important and primary 
$e^+e^-$ collisions are accompanied by several $\gamma \gamma \rightarrow 
{\mathrm{hadrons}}$ interactions. Being mostly confined in the forward regions, this 
$\gamma \gamma$ background reduces the polar angle acceptance for quark flavour tagging 
and dilutes the jet charge separation using jet charge techniques. These experimental 
conditions require efficient and robust algorithms to ensure sensitivity to 
flavour-specific $f \bar f$ production. The statistical accuracies for the determination 
of $\sigma_{f \bar f}$ and $A_{FB}^{f \bar f}$ have been studied using a realistic 
simulation. $b \bar b$ final states have been identified based on the sampling of the 
decay charged multiplicity of the highly boosted $b$ hadrons at {\sc Clic} 
energies~\cite{Battaglia:2000iw}. Similarly to {\sc Lep} analyses, the forward-backward 
asymmetry for $b \bar b$ has been extracted from a fit to the flow of the jet charge 
$Q^{jet}$ defined as $Q^{jet} = \frac{\sum_i q_i |\vec p_i\cdot \vec  T|^k}
{\sum_i |\vec p_i \cdot \vec T|^k}$, where $q_i$ is the
particle charge, $\vec p_i$ its momentum, $\vec T$ the jet thrust unit vector,
$k$  a positive number and the sum is extended to all the particles in a given jet. 
Here the presence of additional particles, from the $\gamma \gamma$ background, causes 
a broadening of the $Q^{jet}$ distribution and thus a dilution of the quark charge 
separation. The track selection and the value of the power parameter $k$ need to be
optimised as a function of the number of overlayed bunch crossings. 
Results for the $e^+e^-\to \bar t t$ channel have been obtained using a dedicated top 
tagging algorithm \cite{laura}. This uses an explicit reconstruction of the $t\to bW$ 
decay and also includes the physics and machine induced backgrounds. For $t \bar t$
forward backward asymmetries the sign of the lepton from the $W^\pm\to\ell^\pm\nu$ 
decay has been used. The results are summarised in terms of the relative statistical 
accuracies $\delta {\cal{O}}/{\cal{O}}$ in Table~\ref{acc}.

However, it is important to stress that at the energy scales considered here, 
electroweak virtual corrections are strongly 
enhanced by Sudakov double logarithms of the type $\log^2 (s/M_W^2)$. Until a complete 
two-loop result will settle the problem, a theoretical error on the cross
section of the order of a percent could be considered \cite{Ciafaloni:2000ey}.
We have not included it in the present analysis.

\TABLE[t]{
\begin{tabular}{|c|c|}
\hline
Observable & Relative Stat. Accuracy \\
           & $\delta {\cal{O}}/{\cal{O}}$ for 1~ab$^{-1}$ \\
\hline \hline
$\sigma_{\mu^+\mu^-}$ & $\pm 0.010$ \\
$\sigma_{b \bar b}$ & $\pm 0.012$ \\
$\sigma_{t \bar t}$ & $\pm 0.014$ \\
$A_{FB}^{\mu\mu}$ & $\pm 0.018$ \\
$A_{FB}^{bb}$ & $\pm 0.055$ \\
$A_{FB}^{tt}$ & $\pm 0.040$ \\
\hline
\end{tabular}
\caption{Relative statistical accuracies on electro-weak
observables, obtained for 1~ab$^{-1}$ of {\sc Clic} data at
$\sqrt{s}$
= 3~TeV, including the effect of $\gamma \gamma \rightarrow
{\mathrm{hadrons}}$ background.}
\label{acc}}

At the LC, the indirect sensitivity to the $Z'$ mass, $M_{Z'}$, can be
parametrised in terms of the available integrated luminosity ${\cal{L}}$, 
and centre-of-mass energy, $\sqrt{s}$. In fact a scaling law for large $M_{Z'}$ 
can be obtained by considering the effect of the $Z'-\gamma$ interference in 
the cross section. For $s<< M_{Z'}^2$ and assuming that the uncertainties
$\delta \sigma$ are statistically dominated, we get the range of mass values giving 
a significant difference from the SM prediction:
\begin{equation}
\frac{|\sigma^{SM} - \sigma^{SM+Z'}|}{\delta \sigma} \propto
\frac{1}{M^2_{Z'}}\sqrt{s{\cal{L}}} > \sqrt{\Delta \chi^2}
\end{equation}
and the sensitivity to the $Z'$ mass scales as:
\begin{equation}
M_{Z'} \propto (s {\cal{L}})^{1/4} \label{resc}
\end{equation}

This relationship shows that there is a direct trade-off possible
between the centre-of-mass energy, $\sqrt{s}$, and the luminosity, ${\cal{L}}$, 
which should be taken into account when optimising the parameters of a high energy 
$e^+e^-$ linear collider.

\FIGURE[t]{
\begin{tabular}{l r}
\includegraphics[width=0.47\textwidth,height=0.5\textwidth]{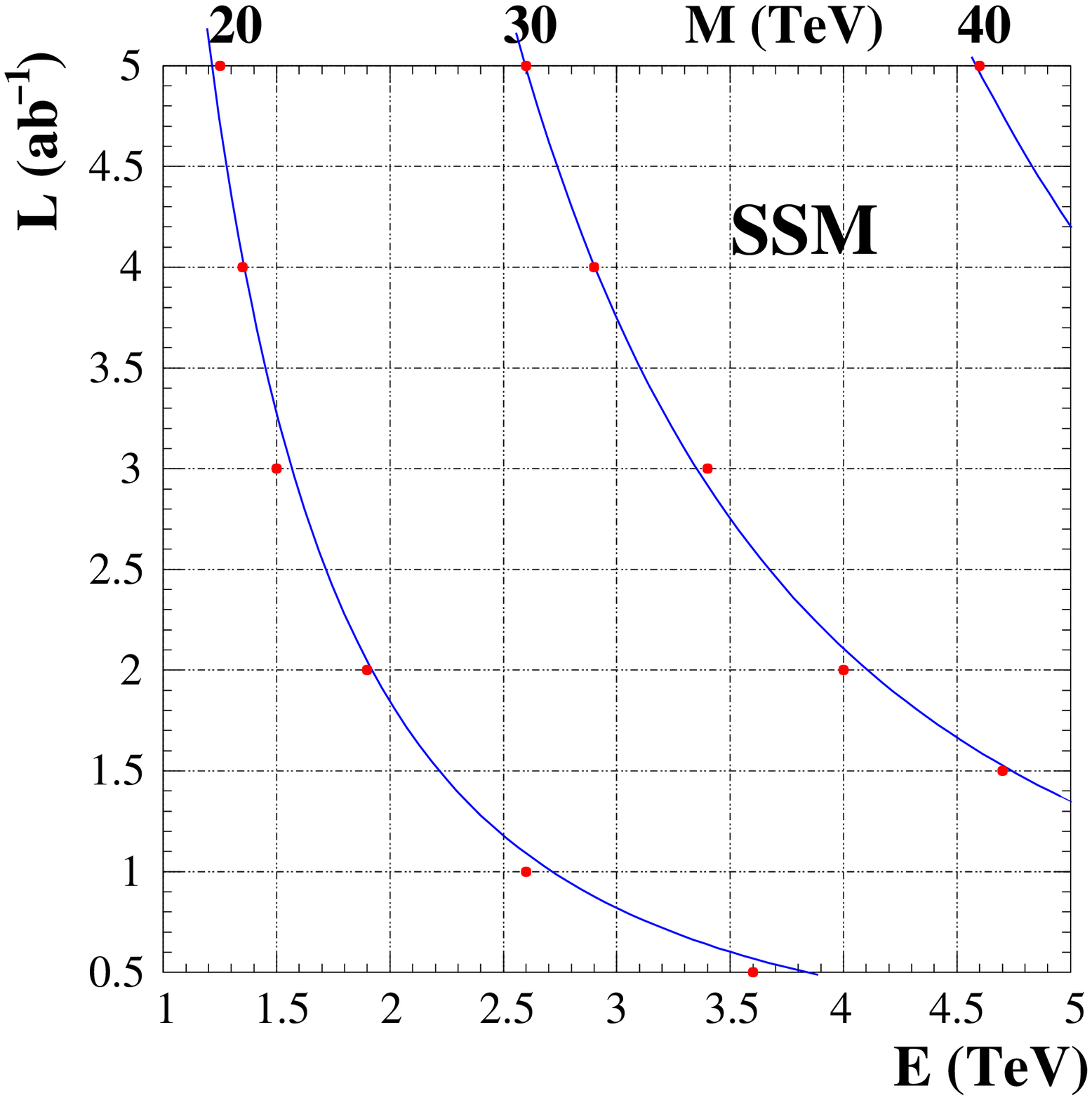}&
\includegraphics[width=0.47\textwidth,height=0.5\textwidth]{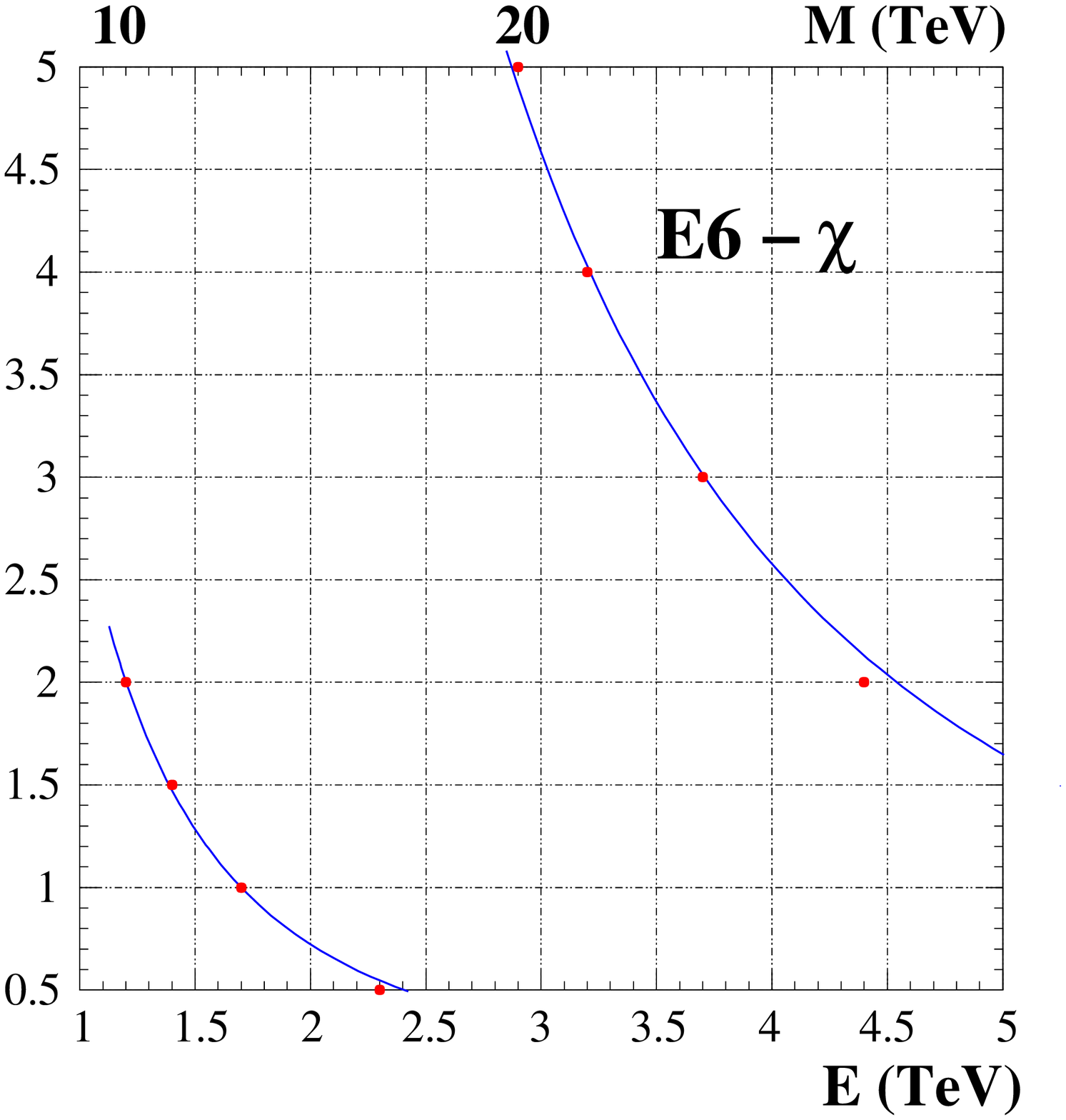}\\
\end{tabular}
\caption{The 95\%~C.L. sensitivity contours in the ${\cal{L}}$ vs.
$\sqrt{s}$ plane for different values of $M_{Z'}$ in the SSM
model (left) and in the $E_6$~$\chi$ model (left).
The points represent the results of the analysis, while curves show the 
behaviour expected from the scaling at eq.~(5.2)
}
\label{fig:zp}}

The $\sigma_{f \bar f}$ and $A_{FB}^{f \bar f}$ ($f = \mu,~b,~t$)
values have been computed, for 1~TeV $< \sqrt{s} <$ 5~TeV, both in the
SM and including the corrections due to the presence of a $Z'$ boson with 
10~TeV $< M_{Z'} <$ 40~TeV, with couplings defined by the models discussed 
in Section~2. Predictions have been obtained by implementing these models 
in the {\sc Comphep} program~\cite{comphep}. Relative
statistical errors on the electroweak observables are obtained by
rescaling the values of Table~\ref{acc} for different energies and
luminosities. The sensitivity has been defined as the largest $Z'$
mass giving a deviation of the actual values of the observables
from their SM predictions corresponding to a SM probability of
less than 5\%.  The SM probability has been defined as the minimum of the global 
probability computed for all the observables and that for each of them, taken 
independently.

This sensitivity has been determined, as a function of the $\sqrt{s}$ energy and 
integrated luminosity ${\cal{L}}$ and compared to the scaling in eq.~(5.2). Results are 
summarised in Figure~\ref{fig:zp}.  For the $\eta$ model the sensitivity is
lower: for example to reach a sensitivity of $M_{Z'}$=20~TeV, more than
10~ab$^{-1}$ of data at $\sqrt{s}$=5~TeV would be necessary.

\FIGURE[t]{
\begin{tabular}{c c}
\includegraphics[width=0.47\textwidth,height=0.5\textwidth]{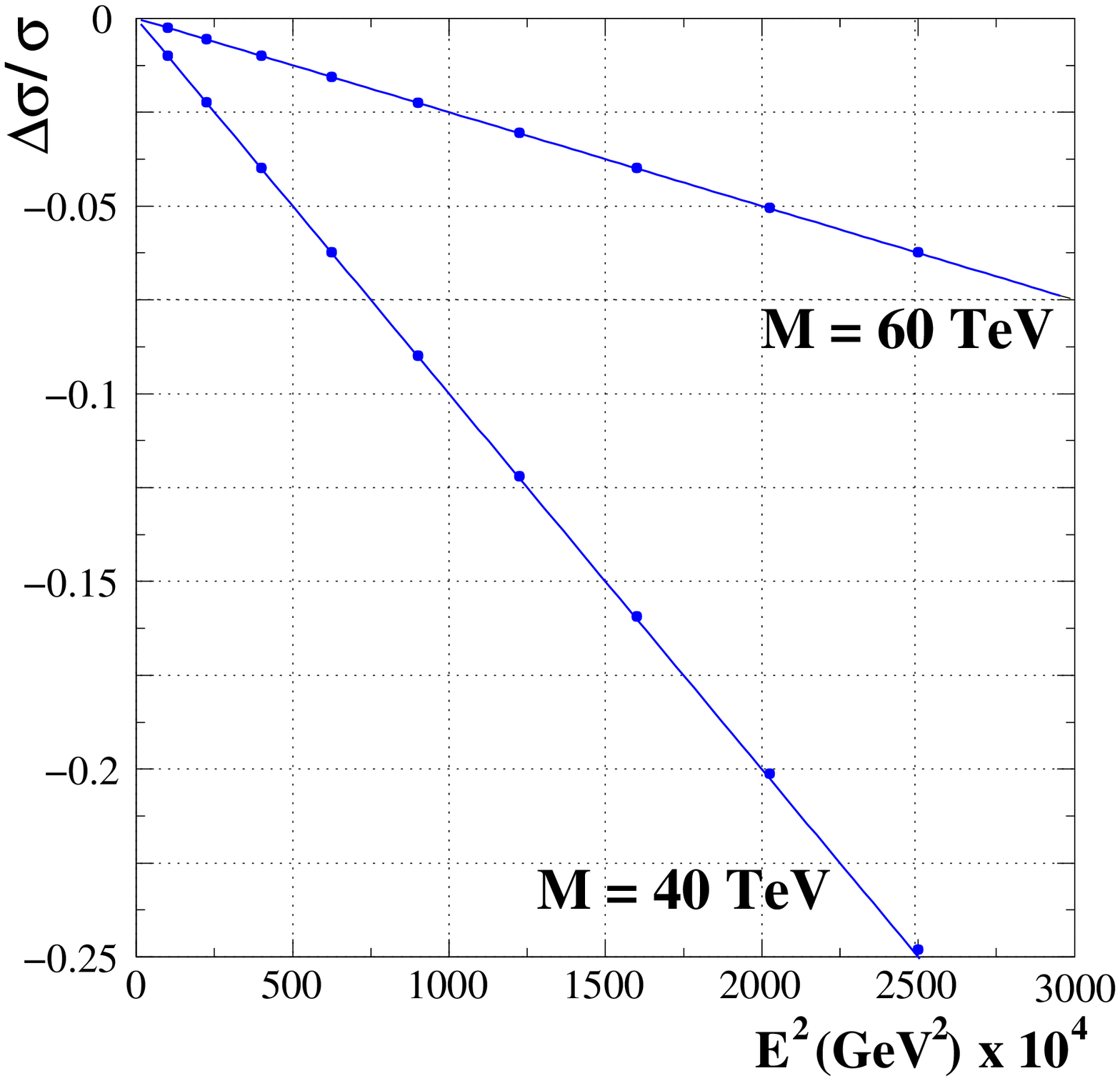}&
\includegraphics[width=0.47\textwidth,height=0.51\textwidth]{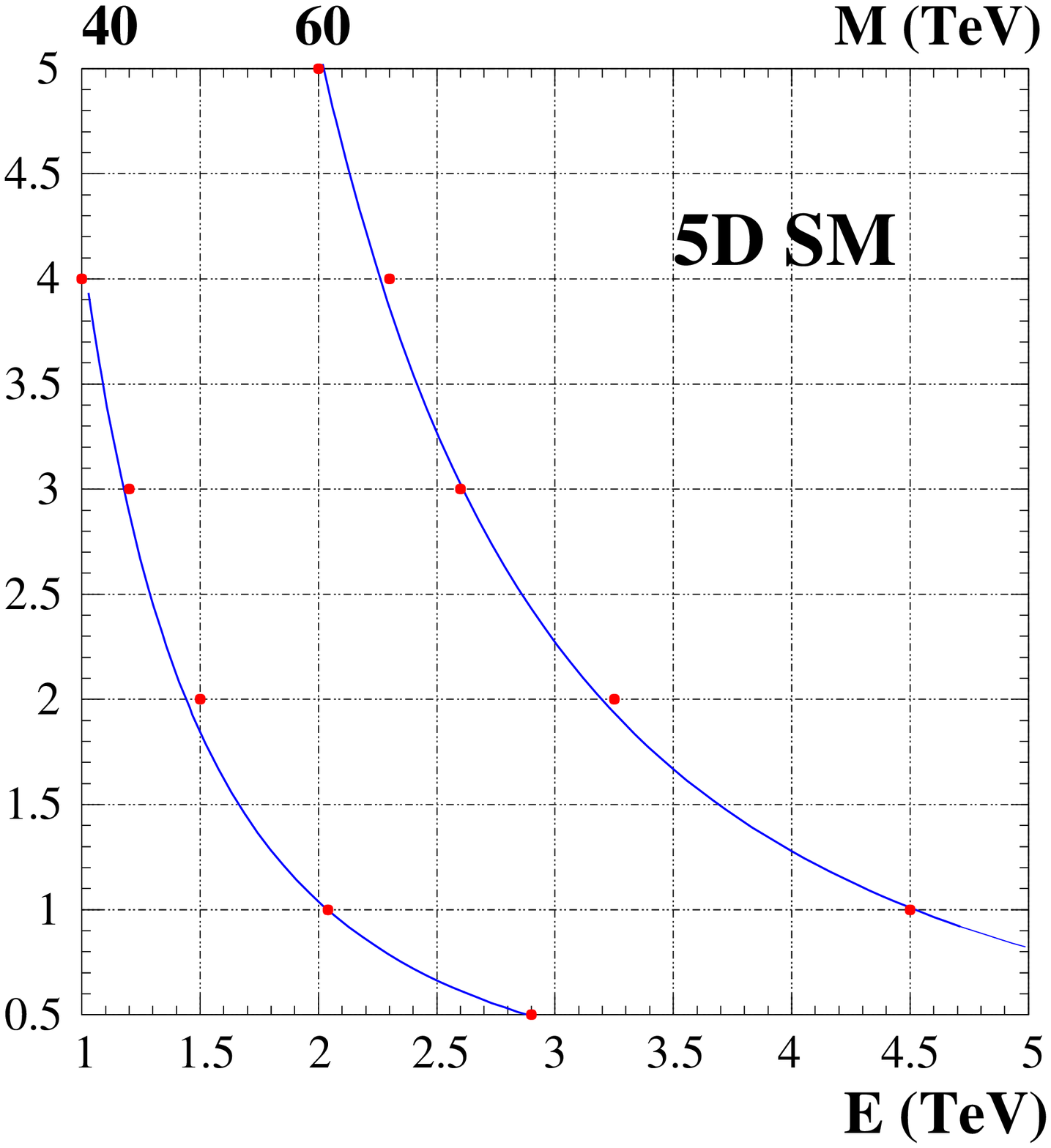}\\
\end{tabular}

\caption{{Left: Scaling of the relative change for the $e^+e^- \to b
\bar{b}$ cross section, for the 5D~SM, as a function of the square of the 
centre-of-mass energy, for two values of the compactification scale $M$.
Right: The 95\%~C.L. sensitivity contours in the ${\cal{L}}$ vs.
$\sqrt{s}$ plane for different values of the compactification scale $M$ in the 
5D~SM. The points represent the results of the analysis, while curves show the 
behaviour expected from the scaling in eq.~(5.2)}
\label{scaling} }
}

In the case of the 5D~SM, we have included only the effect of the 
exchange of the first KK excitations $Z^{(1)}$ and $\gamma^{(1)}$, 
neglecting that of the remaining excitations of the towers, which 
give only small corrections. The scaling law for the limit on $M$ can 
be obtained by considering the interference of the two new nearly degenerate
gauge bosons with the photon in the cross section and taking the
$s<<M^2$ limit. The result is the same as eq.~(\ref{resc}).
The analysis closely follows that for the $Z'$ boson discussed above.
In Figure~ \ref{scaling} we give the sensitivity contours as a
function of $\sqrt{s}$  for different values of $M$. We conclude
that the sensitivity achievable on the compactification scale $M$
for an integrated luminosity of 1~ab$^{-1}$ in $e^+e^-$ collisions
at $\sqrt{s}$ = 3-5~TeV is of the order of 40-60~TeV.
Results for a similar analysis, including all electro-weak
observables, are discussed in~\cite{Rizzo:2001gk}.

\FIGURE[t]{\epsfig{file=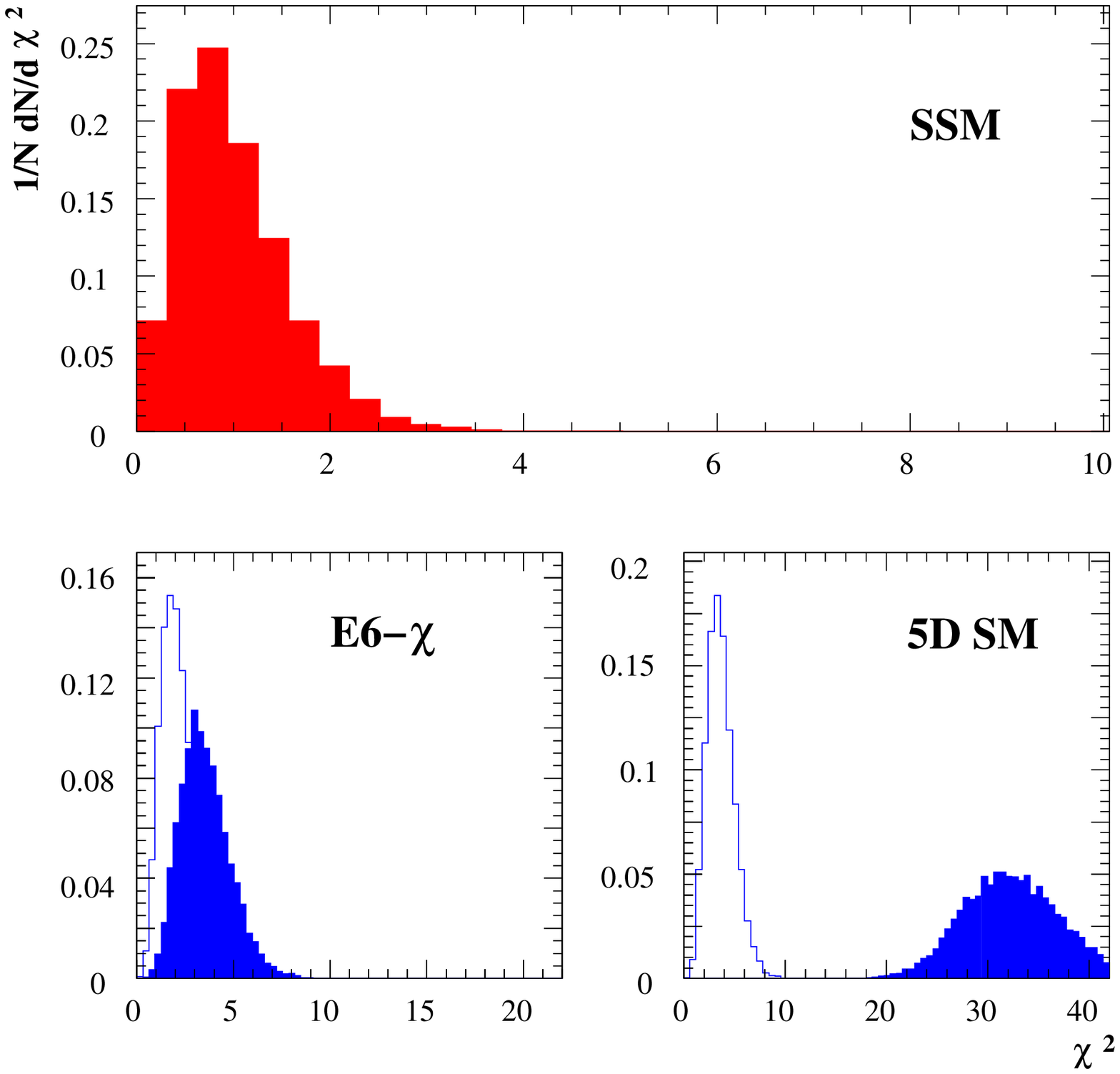,width=12.cm,height=9.cm,clip}
\caption{$\chi^2$ distributions obtained for a set of
pseudo-experiments where the SSM is realised with a $M_{Z'}$ mass of 
20~TeV (upper plot). The corresponding distributions for the $E(6)$ $\chi$
and 5D~SM for the same mass scale (full histograms) and for $M$=40~TeV 
are also shown for comparison in the lower panels. By integrating these
distributions, the confidence levels for discriminating between these 
models, discussed in the text, are obtained. }
\label{fig:sep}}

An important issue concerns the ability to probe the models, once a significant 
discrepancy from the SM predictions would be observed. Since the model parameters 
and the mass scale are {\it a priori} arbitrary, an unambiguous identification of the 
scenario realised is difficult. However, some informations can be extracted by testing 
the compatibility of different models while varying the mass scale.  
Figure~\ref{fig:sep} shows an example of such tests. Taking $M$=20~TeV, ${\cal{L}}$ = 
1~ab$^{-1}$ of {\sc Clic} data at $\sqrt{s}$=3~TeV could distinguish the SSM model from 
the $E_6$~$\chi$ model at the 86\% C.L. and from the 5D~SM at the 99\% C.L.  For a mass 
scale of 40~TeV, ${\cal{L}}$ = 3~ab$^{-1}$ of {\sc Clic} data at $\sqrt{s}$=5~TeV, 
the corresponding confidence levels become 91\% and 99\% respectively.
Further sensitivity to the nature of the gauge bosons could be obtained by 
studying the polarised forward-backward asymmetry $A_{FB}^{pol}$ and the 
left-right asymmetry $A_{LR}$ colliding polarised beams.

\section{Conclusions}

New neutral vector gauge bosons characterise several extensions of the Standard 
Model and may represent the main phenomenology beyond 1~TeV. Their existence and 
properties can be precisely studied at a multi-TeV $e^+e^-$ collider. Present 
bounds are derived from precision electro-weak data and typically constrain the 
masses of these new bosons to be heavier than 1~TeV.  At these scales, they may be 
first observed at the {\sc Lhc} and subsequently studied at {\sc Clic}. Accuracies 
achievable for the determination of their fundamental properties are discussed for 
different classes of models, using realistic assumptions for the experimental 
conditions at {\sc Clic}. Even beyond the kinematical reach for s-channel production, 
a multi-TeV $e^+e^-$ collider could probe the existence of new vector resonances up to 
scales of several tens of TeV by studying the unpolarised electroweak observables. 
Some information regarding the nature of these new resonances could still be gained 
and further sensitivity would be provided by the use of polarised beams.

During the completion of this work two papers have confirmed the results
on the relevance of the QED self energy and vertex corrections in
the calculation of atomic parity violation
\cite{Kuchiev:2002qe, Milstein:2002ai}.

It is a pleasure to thank R.~Casalbuoni, A.~De~Roeck, J.~Hewett, S.~Riemann, T.~Rizzo, 
L.~Salmi and D.~Schulte for discussion and suggestions on several of the topics 
presented in this paper.


\begin{thebibliography}{999}
\def \pra#1#2#3{{\it{Phys.~Rev.}}~A {\bf#1}, #2 (#3)}

\def \prl#1#2#3{{\it{Phys.~Rev.~Lett.}}~{\bf#1}, #2 (#3)}

\bibitem{clic}
{\it A 3~TeV $e^+e^-$ Linear Collider Based on {\sc Clic} Technology},
G. Guignard (editor), CERN-2000-008.

\bibitem{dbess}
R. Casalbuoni, A. Deandrea, S. De Curtis, D. Dominici, F. Feruglio, R.
Gatto and M. Grazzini, Phys.\ Lett.\ {\bf B349} (1995) 533; R.
Casalbuoni, A. Deandrea, S. De Curtis, D. Dominici, R. Gatto and M.
Grazzini, Phys.\ Rev.\ {\bf D53} (1996) 5201.

\bibitem{Wacker:2002ar}
J.~G.~Wacker,
[arXiv:hep-ph/0208235].

\bibitem{Abe:1997fd}
F.~Abe {\it et al.}  [CDF Collaboration],
Phys.\ Rev.\ Lett.\  {\bf 79} (1997) 2192.

\bibitem{kobel}
M. Kobel, talk given at the 31st Int. Conference on High
Energy Physics, Amsterdam 2002.

\bibitem{Langacker:2001ij}
P.~Langacker,
in {\it Proc. of the APS/DPF/DPB Summer Study on the Future of Particle Physics 
(Snowmass 2001) } ed. N.~Graf, [arXiv:hep-ph/0110129].

\bibitem{lep}
LEPEWWG $f \bar f$ Subgroup, LEP2FF/01-02

\bibitem{Casalbuoni:1999yy}
R.~Casalbuoni, S.~De Curtis, D.~Dominici and R.~Gatto,
Phys.\ Lett.\ B {\bf 460} (1999) 135
[arXiv:hep-ph/9905568].

\bibitem{Bennett:1999pd}
S.~C.~Bennett and C.~E.~Wieman,
Phys.\ Rev.\ Lett.\  {\bf 82} (1999) 2484
[arXiv:hep-ex/9903022].

\bibitem{breit}
 A. Derevianko, Phys.\ Rev.\ Lett. {\bf 85} (2000) 1618;
 M. G. Kozlov, S. G. Porsev, and I. I. Tupitsyn, Phys.\ Rev.\ Lett.\ {\bf86} (2001) 3260;
 V.A. Dzuba, C. Harabati, W.R. Johnson and M.S. Safronova, Phys.\ Rev.\ {\bf A63} (2001)
044103.

\bibitem{radapv}
 A.I. Milstein and O.P. Sushkov, [arXiv:hep-ph/0109257]; W.R. Johnson, I. Bednyakov
 and G. Soff, Phys.\ Rev.\ Lett.\ {\bf 87} (2001) 233001.
 
\bibitem{dzuba}
V.A. Dzuba, V.V. Flambaum and J.S.M. Ginges, [arXiv:hep-ph/0204134].

\bibitem{Kuchiev:2002fg}
M.~Y.~Kuchiev and V.~V.~Flambaum,
[arXiv:hep-ph/0206124].

\bibitem{PDG}
J. Erler and P. Langacker, in PDG WWW pages (URL:http://pdg.lbl. gov/)

\bibitem{Godfrey:2002tn}
S.~Godfrey,
in {\it Proc. of the APS/DPF/DPB Summer Study on the Future of Particle 
Physics (Snowmass 2001) } ed. N.~Graf, [arXiv:hep-ph/0201093].


\bibitem{Schulte:2001kh}
D.~Schulte,
CERN-PS-2001-002-AE
{\it Prepared for 5th International Linear Collider Workshop (LCWS 2000), 
Fermilab, Batavia, Illinois, 24-28 Oct 2000}.


\bibitem{peskin}
K.~Yokoya and P.~Chen, in Proc. of the {\it  1989 Particle Accelerator
Conference}, F.~Bennet and L.~Taylor (eds.), IEEE 1989 and
M.~Peskin, LCC Note~0010.


\bibitem{alta}
See for example G. Altarelli  {\it et al.}, Mod.\ Phys.\ Lett.\
{\bf A5} (1990) 495 and Nucl.\ Phys.\ {\bf B342} (1990)  15.


\bibitem{Battaglia:2001fr}
M.~Battaglia, S.~De Curtis, D.~Dominici and S.~Riemann,
in {\it Proc. of the APS/DPF/DPB Summer Study on the Future of Particle Physics 
(Snowmass 2001) } ed. N.~Graf, [arXiv:hep-ph/0112270].


\bibitem{Battaglia:2001dg}
M.~Battaglia, S.~Jadach and D.~Bardin,
in {\it Proc. of the APS/DPF/DPB Summer Study on the Future of Particle Physics 
(Snowmass 2001) } ed. N.~Graf, SNOWMASS-2001-E3015

\bibitem{bess}
R.~Casalbuoni, S.~De Curtis, D.~Dominici and R.~Gatto, 
Phys.\ Lett.\ {\bf B155} (1985) 95; {\it eadem}, Nucl.\ Phys.\ {\bf B282} (1987) 235.

\bibitem{epsi}
G. Altarelli, F. Caravaglios, G.F. Giudice, P. Gambino and  G.
Ridolfi, JHEP {\bf 0106} (2001) 018.

\bibitem{epsilon}
G. Altarelli, [arXiv:hep-ph/0011078].

\bibitem{redi}
R. Casalbuoni, S.~De Curtis. and  M.~Redi, Eur.\ Phys.\ J.\ {\bf C18} (2000) 65.

\bibitem{Casalbuoni:1999mm}
R.~Casalbuoni, A.~Deandrea, S.~De Curtis, D.~Dominici, R.~Gatto and J.~F.~Gunion,
JHEP {\bf 9908} (1999) 011, [arXiv:hep-ph/9904268].

\bibitem{HLZ}  G.F. Giudice, R. Rattazzi and J.D. Wells,
Nucl.\ Phys.\ {\bf B544} (1999) 3; E. A. Mirabelli, M. Perelstein
and M. E. Peskin, Phys.\ Rev.\ Lett.\ {\bf 82} (1999) 2236;
 T. Han, J.D. Lykken, R.-J. Zhang,
Phys.\ Rev.\ {\bf D59} (1999) 105006.

\bibitem{antoniadis} See, for example,
I. Antoniadis, K. Benakli and  M. Quiros, Phys.\ Lett.\ {\bf  B331}
(1994) 313, ibidem {\bf B460} (1999) 176;  T. Rizzo, Phys.\ Rev.\
{\bf D64} (2001) 015003.

\bibitem{Pomarol:1998sd}
A.~Pomarol and M.~Quiros,
Phys.\ Lett.\ {\bf B438} (1998) 255
[arXiv:hep-ph/9806263].

\bibitem{ewlimits}
See, for example,  T.G.~Rizzo and J.D.~Wells, Phys.\ Rev.\ {\bf D61} (2000)
016007; P. Nath and M. Yamaguchi, Phys.\ Rev.\ {\bf  D60} (1999)
116006; M. Masip and A. Pomarol, Phys.\ Rev.\ {\bf D60} (1999)
096005;  A.~Strumia, Phys.\ Lett.\ {\bf B466} (1999) 107.

\bibitem{Casalbuoni:1999ns}
R.~Casalbuoni, S.~De Curtis, D.~Dominici and R.~Gatto,
Phys.\ Lett.\ {\bf B462} (1999) 48
[arXiv:hep-ph/9907355].

\bibitem{Antoniadis:1999bq}
I.~Antoniadis, K.~Benakli and M.~Quiros,
Phys.\ Lett.\  {\bf B460} (1999) 176
[arXiv:hep-ph/9905311].

\bibitem{Battaglia:2000iw}
M.~Battaglia,  in {\it Physics and Experiments with Future Linear
$e^+e^-$ Colliders}, (A.~Para and H.E.~Fisk editors), AIP Conference
Proceedings, New York, 2001, 813, [arXiv:hep-ex/0011099].


\bibitem{laura}
L. Salmi, in Proc. of the {\it Workshop on Physics and Experiments with 
Future Electron-Positron Linear Colliders}, Jeju Island, Korea.

\bibitem{Ciafaloni:2000ey}
P.~Ciafaloni,
[arXiv:hep-ph/0005277] and references therein.

\bibitem{comphep}
A.~Pukhov {\it et al.},
INP-MSU-98-41-542, [arXiv:hep-ph/9908288].


\bibitem{Rizzo:2001gk}
T.~G.~Rizzo,
in {\it Proc. of the APS/DPF/DPB Summer Study on the Future of Particle Physics 
(Snowmass 2001) } ed. N.~Graf, [arXiv:hep-ph/0108235].

\bibitem{Kuchiev:2002qe}
M.~Y.~Kuchiev and V.~V.~Flambaum,
[arXiv:hep-ph/0209052].

\bibitem{Milstein:2002ai}
A.~I.~Milstein, O.~P.~Sushkov and I.~S.~Terekhov,
[arXiv:hep-ph/0208227].

\end{thebibliography}
\end{document}